\documentclass[10pt,superscriptaddress,aps,prx,twocolumn,showpacs,nofootinbib, longbibliography, notitlepage,floatfix]{revtex4-2}
\usepackage{etex}
\usepackage{amsmath,amssymb,amsthm,bm}
\usepackage[breaklinks=true,colorlinks,citecolor=blue,linkcolor=blue,urlcolor=blue]{hyperref}
\usepackage[pdftex]{graphicx}
\usepackage{braket}
\usepackage{comment}
\usepackage{physics}
\usepackage{color}
\usepackage{natbib}
\usepackage{amsmath,blkarray}
\usepackage{mathtools}
\usepackage[normalem]{ulem}
\usepackage{latexsym}
\usepackage{tabularx, booktabs}
\usepackage{graphics,epstopdf}
\usepackage{graphicx}
\usepackage{float}
\usepackage{amsfonts}
\usepackage{tikz}
\usepackage{quantikz}
\usepackage{color,soul}

\newcommand{\be}{\begin{equation}}
\newcommand{\ee}{\end{equation}}
\newcommand{\ba}{\begin{eqnarray}}
\newcommand{\ea}{\end{eqnarray}}

\usepackage{multirow}
\usepackage{orcidlink}
\usepackage{url}

\usepackage{colortbl}
\usepackage{lettrine}
\usepackage{setspace}

\usepackage{wrapfig}
\newcommand{\eqlabel}[1]{equation~\eqref{#1}}
\newcommand{\figlabel}[1]{Fig.~\ref{#1}}

\makeatletter
\def\bibsection{%
   \par
   \begingroup
    \baselineskip26\p@
    \bib@device{\hsize}{72\p@}%
   \endgroup
   \nobreak\@nobreaktrue
   \addvspace{19\p@}%
  }%
\makeatother

\begin{document}
\title{Digitized counterdiabatic quantum critical dynamics}

\author{Anne-Maria Visuri$^{\orcidlink{0000-0002-4167-7769}}$}
\email{annemariavisuri@gmail.com}
\affiliation{Kipu Quantum GmbH, Greifswalderstrasse 212, 10405 Berlin, Germany}

\author{Alejandro Gomez Cadavid$^{\orcidlink{0000-0003-3271-4684}}$}
\affiliation{Kipu Quantum GmbH, Greifswalderstrasse 212, 10405 Berlin, Germany}
\affiliation{Department of Physical Chemistry, University of the Basque Country EHU, Apartado 644, 48080 Bilbao, Spain}

\author{Balaganchi A. Bhargava$^{\orcidlink{0009-0006-7595-8776}}$} 
\affiliation{Kipu Quantum GmbH, Greifswalderstrasse 212, 10405 Berlin, Germany}

\author{Sebastián V. Romero$^{\orcidlink{0000-0002-4675-4452}}$}
\affiliation{Kipu Quantum GmbH, Greifswalderstrasse 212, 10405 Berlin, Germany}
\affiliation{Department of Physical Chemistry, University of the Basque Country EHU, Apartado 644, 48080 Bilbao, Spain}

\author{András Grabarits$^{\orcidlink{https://orcid.org/0000-0002-0633-7195}}$}
\affiliation{Department of Physics and Materials Science, University of Luxembourg, L-1511 Luxembourg, Luxembourg}

\author{Pranav Chandarana$^{\orcidlink{0000-0002-3890-1862}}$}
\affiliation{Kipu Quantum GmbH, Greifswalderstrasse 212, 10405 Berlin, Germany}
\affiliation{Department of Physical Chemistry, University of the Basque Country EHU, Apartado 644, 48080 Bilbao, Spain}

\author{Enrique Solano$^{\orcidlink{0000-0002-8602-1181}}$}
\email{enr.solano@gmail.com}
\affiliation{Kipu Quantum GmbH, Greifswalderstrasse 212, 10405 Berlin, Germany}

\author{Adolfo del Campo$^{\orcidlink{ 0000-0003-2219-2851}}$}\email{adolfo.delcampo@uni.lu}
\affiliation{Department of Physics and Materials Science, University of Luxembourg, L-1511 Luxembourg, Luxembourg}
\affiliation{Donostia International Physics Center, E-20018 San Sebastian, Spain}

\author{Narendra N. Hegade$^{\orcidlink{0000-0002-9673-2833}}$}
\email{narendrahegade5@gmail.com}
\affiliation{Kipu Quantum GmbH, Greifswalderstrasse 212, 10405 Berlin, Germany}

\begin{abstract}
We experimentally demonstrate that a digitized counterdiabatic quantum protocol reduces the number of topological defects created during a fast quench across a quantum phase transition. To show this, we perform quantum simulations of one- and two-dimensional transverse-field Ising models driven from the paramagnetic to the ferromagnetic phase. We utilize superconducting cloud-based quantum processors with up to 156 qubits. Our data reveal that the digitized counterdiabatic protocol reduces defect formation by up to 48\% in the fast-quench regime—an improvement hard to achieve through digitized quantum annealing under current noise levels. The experimental results closely match theoretical and numerical predictions at short evolution times before deviating at longer times due to hardware noise. In one dimension, we derive an analytic solution for the defect number distribution in the fast-quench limit. For two-dimensional geometries, where analytical solutions are unknown and numerical simulations are challenging, we use advanced matrix product state methods. Our findings indicate a practical way to control topological defect formation during fast quenches and highlight the utility of counterdiabatic protocols for quantum optimization and quantum simulation in material design on current quantum processors.
\end{abstract}

\maketitle

The critical dynamics close to phase transitions has universal features that connect phenomena at vastly different energy scales: the polarization of magnetic materials, the superfluid transition of liquid helium, and the inflationary dynamics of the early universe~\cite{zurek1996cosmological}. Close to a continuous symmetry-breaking phase transition, the spatial correlation length and relaxation time diverge. 
As a result, a system driven through the phase transition at a nonzero rate fails to reach its instantaneous equilibrium with a uniform phase. Instead, domains of different broken-symmetry states are formed, leading to the excitation of topological defects such as domain walls and vortices.

For slow driving protocols, the Kibble-Zurek mechanism (KZM) provides a universal framework for describing this process: the density of defects scales as a power law of the driving rate, with an exponent determined by the universality class of the phase transition~\cite{delcampo2014universality}. The mechanism, originally conceived for continuous phase transitions, was recently extended to phase transitions with tunable order~\cite{Rams2019,suzuki2024topological,Sadhasivam2024}. Its applicability to classical and quantum phase transitions~
has been verified in numerous theoretical and experimental studies, and
recent quantum simulation experiments with Rydberg atoms~\cite{keesling2019quantum}
and superconducting circuits~\cite{weinberg2020scaling, bando2020probing, king2022coherent, miessen2024benchmarking, andersen2024thermalization,teplitskiy2024statistics} have demonstrated the quantum Kibble-Zurek mechanism (QKZM) in new regimes. While most demonstrations have realized the one-dimensional (1D) transverse-field Ising model (TFIM), recent work has also explored two-dimensional (2D) interacting systems \cite{ebadi2021quantum, king2024computational} for which classical simulations are difficult \cite{schmitt2022quantum}. Connections between the phenomenology of QKZM and quantum speed limits have deepened its theoretical foundations \cite{puebla2020kibble, carolan2022counterdiabatic}.
However, the QKZM applies only in the near-adiabatic regime, when the defect density is small. The power-law scaling with quench rate breaks down in fast quenches, where the defect density saturates into a plateau that depends on the quench depth rather than rate~\cite{xia2021kibblezurekmechanismrapidly,zeng2023universal,grabarits2025drivingquantumphasetransition}.

\begin{figure*}[!tb]
    \centering
    \includegraphics[width=\linewidth]{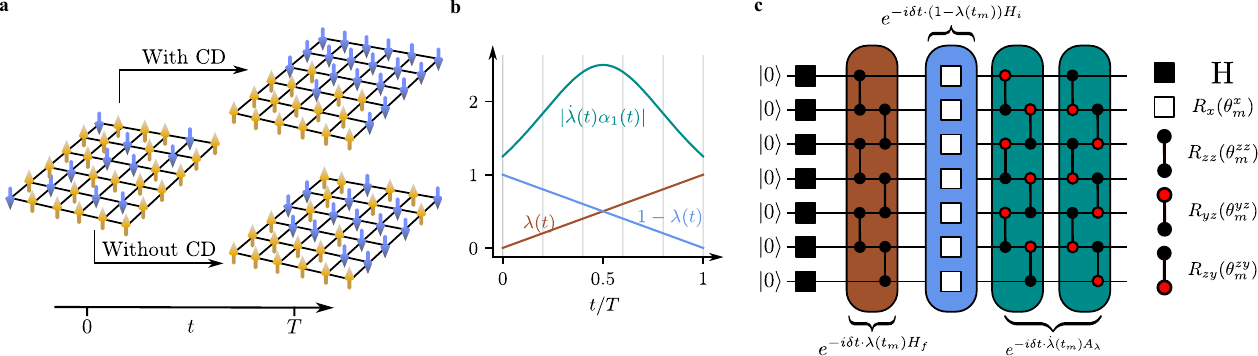}
    \caption{\textbf{A schematic illustration of the initial and final states resulting from CD-assisted evolution and digitized annealing without CD.} \textbf{a,} The spin system is driven across a phase transition from the paramagnetic to the ferromagnetic phase. Counterdiabatic evolution results in fewer kinks in the magnetization in the final state at time $t = T$. \textbf{b,} The time-dependent factors in the Hamiltonian $\mathcal{H}(\lambda) = H(\lambda) + \dot{\lambda} A_{\lambda}$. The magnitude of the coefficient $|\dot{\lambda}(t) \alpha_1(t)|$ of the CD Hamiltonian is largest at the critical point where excitations have the lowest energy cost and are most likely to occur. In one dimension, the critical point $g = J$ is crossed at $t/T = 0.5$.
    The scheduling function $\lambda(t) = t/T$ is chosen as linear, and the vertical lines indicate that time is discretized into steps of size $\delta t$. \textbf{c,} The circuit that implements the CD evolution.
 The colored boxes correspond to a single time step with $t_m = m \delta t$, and omitting the green boxes results in the implementation of digitized annealing. The black squares denote Hadamard gates, and $R_x(\theta)=\exp(-i\frac{\theta}{2} X)$, $R_{zz}(\theta)=\exp(-i\frac{\theta}{2} Z \otimes Z)$, $R_{yz}(\theta)=\exp(-i\frac{\theta}{2} Y \otimes Z)$, and $R_{zy}(\theta)=\exp(-i\frac{\theta}{2} Z \otimes Y)$ are single- and two-qubit gates with the Pauli matrices $X$, $Y$, and $Z$.}
     \label{fig:schematic}
\end{figure*}

The QKZM and its extensions not only provide a universal description of nonequilibrium critical phenomena but are also relevant in the practical pursuit of quantum computing relying on the adiabatic theorem. 
For systems with a finite energy gap above the ground state, a driving rate below the gap is known to preserve adiabaticity. At a quantum phase transition, however, the gap closes in the thermodynamic limit and excitations become inevitable. Excitations along the adiabatic path are detrimental in applications such as quantum optimization~\cite{abbas2024challenges} and quantum state preparation~\cite{clinton2024towards,maskara2025programmable}, where they reduce the fidelity of the final state~\cite{suzuki2011kibble}.

In adiabatic quantum computing, 
an easy-to-prepare ground state of an initial
Hamiltonian is evolved adiabatically to that of a final
Hamiltonian encoding the solution~\cite{albash2018adiabatic}. 
When the Hamiltonian parameters are changed sufficiently slowly, the system remains in its instantaneous ground state. 
However, the time scale required for adiabaticity typically exceeds the coherence time of current and near-term quantum computers, making adiabatic quantum computing infeasible. The Kibble-Zurek scaling governing the slow crossing of a quantum critical point explicitly reflects these challenges, given that the power-law dependence of the defect density on the driving time scale, the quench time $T$, is governed by an exponent that generally takes fractional values smaller than $1$. For instance, for a 1D Ising chain, the defect density scales as $1/T^{1/2}$ \cite{zurek2005dynamics,dziarmaga2005dynamics,polkovnikov2005universal}: reducing it by a factor of 10 requires quench times 100 times longer. It is thus necessary to find driving schemes that circumvent the requirement of slow driving for defect suppression.

Control protocols called shortcuts to adiabaticity~\cite{guery-odelin2019shortcuts} offer a solution to this problem, allowing the evolution time to be shortened at the cost of additional control fields.
One promising technique in this category is counterdiabatic driving (CD), where additional terms added to the Hamiltonian exactly cancel out transitions to excited states \cite{demirplak2003adiabatic,berry2009transitionless}. This approach has been successfully applied to a variety of problems~\cite{delcampo2012assisted, saberi2014adiabatic,hegade2021shortcuts, hegade2022digitized, chandarana2022digitized, chandarana2023digitized}.
Beyond accelerating adiabatic dynamics, CD has been proposed
to reduce the excitations created at critical points where adiabaticity cannot be maintained~\cite{delcampo2012assisted,saberi2014adiabatic}. To our knowledge, however, its effect on topological defects in many-body systems has remained untested
experimentally.

In this work, we provide the first experimental evidence that CD suppresses the formation of topological defects in fast quenches across quantum phase transitions. Using IBM quantum processors, we demonstrate a reduced defect density compared to digitized annealing without CD (see~\figlabel{fig:schematic}a). We further characterize the breakdown of the Kibble-Zurek scaling and the saturation of the defect density in fast quenches experimentally and theoretically, both with and without CD.
Our results establish CD protocols as a promising tool for improving the accuracy of quantum optimization, quantum simulation, and quantum state preparation.

\section*{Results}

\subsection*{Breakdown of Kibble-Zurek scaling in fast quenches}

The KZM description of critical dynamics identifies a freeze-out time measured with respect to the time at which the transition occurs. In the frozen regime, the system cannot reach its instantaneous equilibrium due to the diverging relaxation time. The correlation length in this regime corresponds to the average size of the broken-symmetry domains formed in the phase transition, and the inverse correlation length sets the average density of defects. According to KZM,
the average density of point-like defects follows a universal scaling law with
the total quench time $T$. 
Recent works beyond the KZM paradigm have shown that not only the average density but also the entire defect number distribution is universal, and all cumulants share the same power-law with~$T$~\cite{delcampo2018,Cui2020,bando2020probing,king2022coherent}.

Deviations from the Kibble-Zurek scaling are known to occur in fast quenches, where the defect density instead reaches a plateau and is independent of the quench time. In this limit, the defect density and the critical quench time were proposed to follow a universal scaling with the quench depth---the distance of the final control parameter from its critical value~\cite{xia2021kibblezurekmechanismrapidly,zeng2023universal}. Scaling laws with system size were also investigated~\cite{uhlmann2010systemsize}, though counterdiabatic dynamics has not been considered in this context. 
Here, we study the reduction of the defect density plateau with CD in different geometries at a fixed quench depth.
As a benchmark for the experiments in one dimension, we derive the exact solution for the defect density distribution in Supplementary Note~4.

\subsection*{Digitized counterdiabatic protocol}

To evolve the ground state of an initial Hamiltonian $H_i$ into that of a final Hamiltonian $H_f$, one can introduce a time-dependent control parameter $\lambda(t)$ in the total Hamiltonian
\begin{equation}
H(\lambda) = \left(1 - \lambda(t) \right) H_i + \lambda(t) H_f.
\label{eq:adiabatic_hamiltonian}
\end{equation}
We choose a linear function $\lambda(t) = t/T$, which changes from $\lambda(0) = 0$ to $\lambda(T) = 1$ at the final time~$T$. If the change is sufficiently slow, the process is adiabatic. 
Specifically, we implement the paramagnetic-to-ferromagnetic phase transition of the TFIM with
\begin{align}
H_i = -g \sum_{j = 1}^{N} \hat{X}_j, 
\hspace{1cm}
H_f = -J \sum_{\langle i,j \rangle} \hat{Z}_i \hat{Z}_j,
\label{eq:initial_and_final_hamiltonian}
\end{align}
where $\hat{X}_i$ and $\hat{Z}_i$ denote the Pauli operators acting on site $i$, $N$ is the number of qubits, and $\langle i,j \rangle$ indicates nearest neighbors. The transverse field is denoted by $g$, and the Ising interaction by $J$, and we adopt natural units with $\hbar=1$. In our experiments, we set $g=J = 1$, and the energy and time scales are given by $J$ and $1/J$, respectively. The initial state is the ground state of $H_i$: $\ket{+}^{\otimes N} = \left[ (\ket{0} + \ket{1})/\sqrt{2} \right]^{\otimes N}$.

Counterdiabatic driving enables the implementation of adiabatic reference trajectories in arbitrarily short times. Diabatic excitations are canceled out by an auxiliary term added to the Hamiltonian, $\mathcal{H}(\lambda) = H(\lambda) + \dot{\lambda} A_{\lambda}$, where $A_{\lambda}$ is the adiabatic gauge potential (AGP)~\cite{demirplak2003adiabatic,berry2009transitionless,kolodrubetz2017geometry}. The time-dependent factors in $\mathcal{H}(\lambda)$ are illustrated in~\figlabel{fig:schematic}b.
Solving the AGP exactly requires diagonalizing the many-body Hamiltonian and is, therefore, only possible for small systems. Furthermore, $A_{\lambda}$ generally consists of highly non-local many-body couplings, which prohibits its exact implementation. Local approximations are therefore often employed~\cite{delcampo2012assisted,saberi2014adiabatic,claeys2019floquet,takahashi2024shortcuts,barone_counterdiabatic2024,morawetz2024efficient}. One such approximation is obtained through a nested-commutator (NC) series expansion~\cite{claeys2019floquet}
\begin{equation}
 A_\lambda^{l} = i\sum_{k=1}^l \alpha_k(t)\hat{O}_{2k-1}(t),
 \label{eq:nested_commutator}
\end{equation}
where $\hat{O}_k(t) = [H, \hat{O}_{k-1}(t)]$ and $\hat{O}_0 = \partial_\lambda H$. The sum is truncated at order $l$, which controls the locality of the operators that are included. The variational parameters $\alpha_k$ are found by minimizing the action $S_{\lambda}(A_{\lambda}) = \text{Tr}\left(G_{\lambda}^{\dagger} G_{\lambda}^{\phantom{\dagger}} \right)$ with $G_{\lambda} = \partial_{\lambda} H + i[A_{\lambda}, H]$ (Supplementary Note~3). 
Using~\eqlabel{eq:adiabatic_hamiltonian}, we obtain the first-order approximation
\begin{equation}\label{eq:cd}
    A_{\lambda}^1 = 2 g J \alpha_1(\lambda) \sum_{\langle i,j \rangle} \left( \hat{Y}_i \hat{Z}_{j} + \hat{Z}_i\hat{Y}_{j}  \right).
\end{equation}
Implementing~\eqlabel{eq:cd} is not straightforward on analog devices, and we perform the quantum simulations in a digitized manner using IBM hardware (Methods). The corresponding quantum circuit is illustrated in~\figlabel{fig:schematic}c.

\subsection*{Quantum simulation}

\begin{figure*}[tb]
    \centering
    \includegraphics[width=\textwidth]{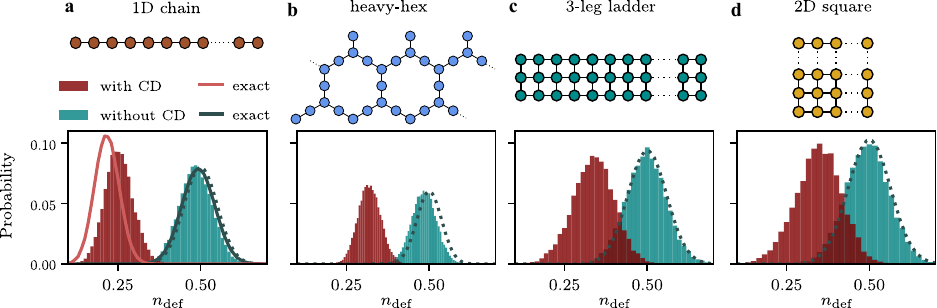}
    \caption{\textbf{Measured distributions of the defect density at the final evolution time $\bm{T = 0.2/J}$.} We consider different geometries: \textbf{a,} a one-dimensional chain of length $N = 100$, \textbf{b,} a 2D heavy-hexagonal lattice of $156$ sites, \textbf{c,} a three-leg ladder of length $N_x = 15$, and \textbf{d,} a square lattice of size $N_x \times N_y = 6 \times 6$. In the presence of CD, the density of defects is reduced on average in all geometries. The histograms correspond to experimental data with 10000 to 20000 samples. The bin width is determined by the number of edges and thus varies. The solid lines in panel \textbf{a} correspond to the exact solution of the 1D transverse-field Ising model (Supplementary Note~4), while for the other geometries, the distributions cannot be computed exactly. The data without CD shows a good agreement with the exact solution. For the counterdiabatic evolution, the data is shifted to larger values due to hardware errors. The dotted lines in each panel are the normal distributions with mean $\kappa_1 = 0.5$ and variance $\kappa_2 = 0.25$ obtained in the initial state $\ket{+}^{\otimes N}$ in the infinite-size limit (Methods).}\label{fig:histograms}
\end{figure*}

\begin{figure*}[!tb]
    \centering
    \includegraphics[width=\linewidth]{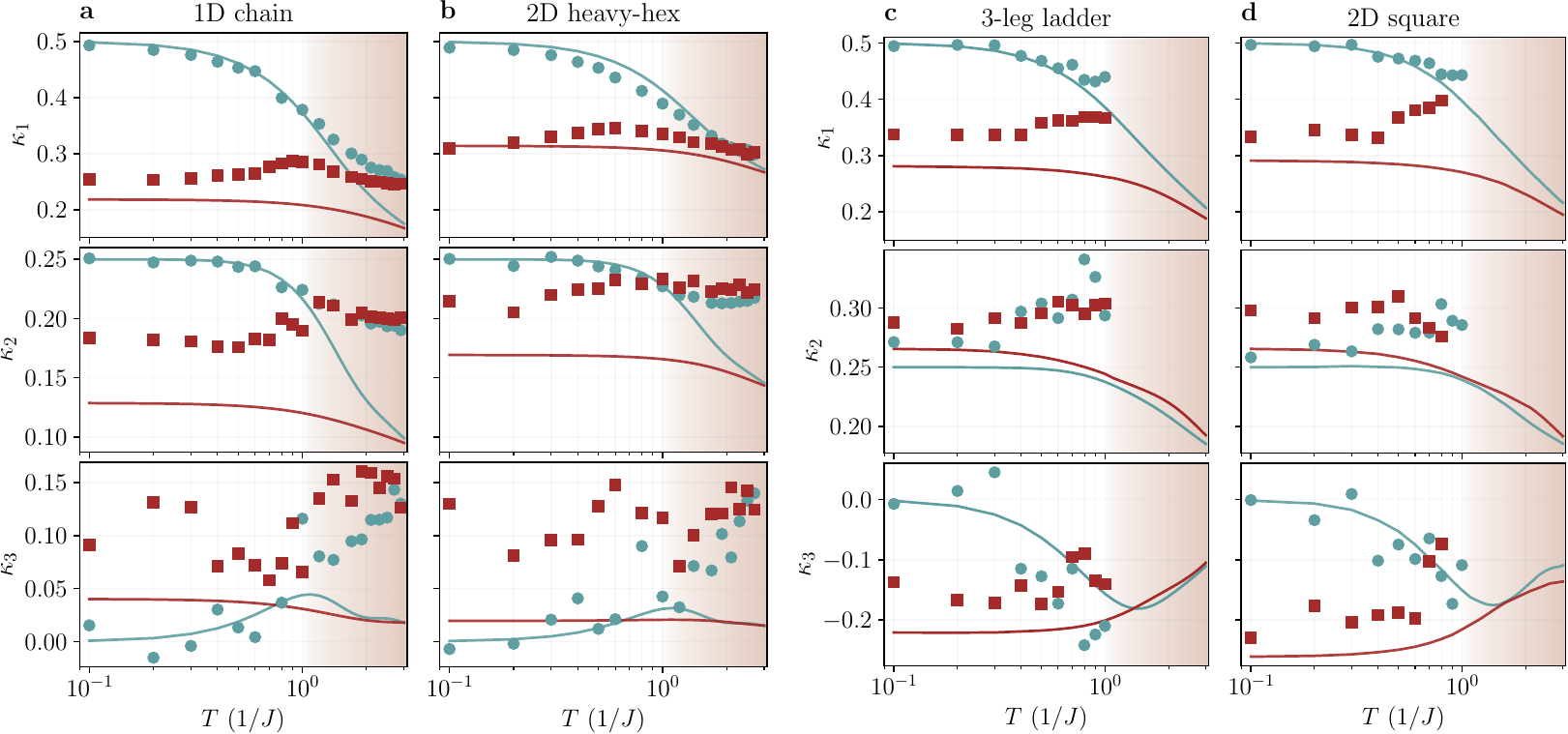}
    \caption{\textbf{Cumulants of the defect density distribution, $\bm{\kappa_1}$, $\bm{\kappa_2}$, and $\bm{\kappa_3}$, as functions of the total evolution time $\bm{T}$ with and without CD.} The geometries are as in~\figlabel{fig:histograms}. The markers correspond to experimental data, and the solid lines correspond to \textbf{a,} the exact solutions (Supplementary Note~4) or \textbf{b-d,} to numerical simulations with MPS (Methods). In all cases, the mean value of the density of defects $\kappa_1$ is reduced by CD. The shaded region indicates the crossover to the KZ scaling regime. \textbf{a, b,} The variance $\kappa_2$ of the defect density is reduced by CD while the skewness of the distribution $\kappa_3$ has a more complex behavior. \textbf{c, d,} The variance is slightly increased by CD, and the skewness has negative values, unlike for the 1D and heavy-hexagonal geometries. Due to increasing errors at large $T$, we only include data up to $T = 1/J$. Each experimental data point is an average of between 10000 and 20000 samples. The standard errors are smaller than the marker size.}
    \label{fig:cumulants}
\end{figure*}

To characterize defect formation, we measure the expectation value of the density of defects $\langle \hat{n}_{\text{def}}\rangle = \frac{1}{2 N_e}\sum_{\langle i,j \rangle}\langle 1 - \hat{Z}_{i} \hat{Z}_{j}\rangle$,
where $N_e$ is the number of edges between qubits. In the Ising model, these defects correspond to kinks in the magnetization where $\langle Z_i \rangle$ changes sign. We further analyze the first three cumulants of the defect density distribution: $\kappa_1 = \langle \hat{n}_{\text{def}} \rangle$, $\kappa_2 = N_e \left\langle \left( \hat{n}_{\text{def}} - \langle \hat{n}_{\text{def}} \rangle \right)^2 \right\rangle$, and $\kappa_3 = N_e^2 \left\langle \left( \hat{n}_{\text{def}} - \langle \hat{n}_{\text{def}} \rangle \right)^3 \right\rangle$.
These correspond, respectively, to the mean, variance, and third central moment, which dictates the skewness of the distribution (Methods).

We perform quantum simulations in the fast-quench regime where the dynamics assisted by CD differs from digitized annealing. We consider various geometries as illustrated in~\figlabel{fig:histograms}. As the simplest geometry, we simulate a 1D spin chain of length $N = 100$ with open boundary conditions, embedded into the heavy-hexagonal qubit layout of \texttt{ibm\_fez}. 
Fig.~\ref{fig:histograms}a shows the corresponding defect density distributions measured at the final evolution time $T = 0.2/J$ with and without CD. 
The two distributions are clearly separated, indicating that the density of defects is suppressed in the presence of CD. The data without CD has a nearly Gaussian distribution, while with CD, the distribution has a small positive skewness, indicating a longer tail above the mean. Fig.~\ref{fig:histograms}a also shows, as solid lines, the Poisson binomial distributions obtained from the exact solution of the TFIM (see Methods and Supplementary Note~4). 
Without CD, the experimental data agrees well with the theoretical prediction. In the presence of CD, the experimental distribution is shifted to larger values and slightly broadened with respect to the theoretical one due to experimental errors (Supplementary Note~2).

In \figlabel{fig:histograms}a, we see that at $T = 0.2/J$, the exact reference distribution without CD is very close to the limiting normal distribution in the initial state, drawn as a dotted line. This means that in such rapid quenches, the system does not have time to evolve towards the ferromagnetic state, but the dynamics freezes almost immediately. This finding is reproduced by the experimental data. In the presence of CD, on the other hand, both the theoretical prediction and the measured mean and variance of the defect density are considerably reduced compared to the initial-state one. The suppression of defects persists for arbitrarily short quench times, where the unitary time evolution operator becomes independent of $T$ (Supplementary Note~4), and it occurs already for the first-order approximation of the CD term. This prediction is supported by the data, measured down to $T = 0.1/J$.

We extend the analysis of fast-quench defect statistics to lattice geometries of varying dimension: a 2D heavy-hexagonal lattice~\cite{chamberland2020topological}, a three-leg ladder, and a 2D square lattice. The corresponding defect density distributions are shown in~\figlabel{fig:histograms}b-d, respectively. For the heavy-hexagonal lattice, the defect density distributions with and without CD are clearly separated, while for the three-leg ladder and the 2D square lattice, the distributions have a larger overlap but still show a reduction of the mean value in the presence of CD. 
Since the TFIM is not exactly solvable in these 2D geometries, we compare the data to the limiting normal distribution in the initial state. Similarly to the 1D case, the distributions without CD are very close to the normal distribution, while applying CD leads to a shift already at very short quench times.

We further analyze the cumulants of the defect density distributions as functions of the total evolution time. The cumulants for the 1D model are shown in~\figlabel{fig:cumulants}a. The mean value $\kappa_1$ has an initial plateau that indicates the breakdown of the QKZM. Although this plateau occurs for both types of dynamics, in the presence of CD, it has a lower value. This reduction of defects is seen both in the experimental data and in the theoretical prediction and occurs already for the first-order NC approximation. We expect higher-order approximations to further reduce the defect density~\cite{delcampo2012assisted}, which we show for the second-order NC approximation in the 1D model in Supplementary Figure~4.
As in~\figlabel{fig:histograms}a, the plateau in the experimental data is shifted with respect to the theoretical one. The reduction of defects with CD persists up to evolution times $T \approx 3/J$, where the curves with and without CD coincide as the CD term $\dot{\lambda} A_{\lambda}$ becomes negligible. 
Without CD, the initial plateau of $\kappa_1$ turns into the power-law decay expected for the KZ scaling around $T \approx 0.5/J$. The experimental data shows a decrease that matches the theoretical prediction and starts to deviate at \(T \gtrsim 1/J\) due to the increasing hardware noise. For CD dynamics, the KZ scaling regime occurs at larger total evolution times. 

The measured $\kappa_2$ and $\kappa_3$ in~\figlabel{fig:cumulants}a agree with the theoretical predictions at short quench times in the absence of CD, while for the CD dynamics, both are shifted to larger values. This shift corresponds to the broadening of the distribution in~\figlabel{fig:histograms}a.

The heavy-hexagonal lattice is implemented by using the full 156-qubit layout of \texttt{ibm\_marrakesh}. 
The cumulants are shown as functions of the total evolution time in~\figlabel{fig:cumulants}b. We obtain reference solutions through time-dependent matrix product states (MPS) simulations (Methods). For the 1D chain, we find that the density of defects saturates into a plateau at short evolution times. Interestingly, the deviation of the experimentally measured defect density from the reference solution is smaller than that for the 1D system, both with and without CD. 
The distributions have a positive skewness, which has a nonmonotonic behavior similar to that observed in the 1D case. Such similarities may be consistent with the low connectivity present in the heavy-hexagonal lattice: 
the degree of connectivity $N_e/N = 6/5 = 1.2$ is closer to the 1D value $N_e/N = 1$ than the one for a 2D square lattice, $N_e/N =2$~\cite{miessen2024benchmarking}. 
The experimental defect density in~\figlabel{fig:cumulants}b, without CD, roughly agrees with the results reported in Ref.~\cite{miessen2024benchmarking} without error mitigation or suppression. 

The three-leg ladder geometry and the 2D square lattice are also embedded into the heavy-hexagonal layout of \texttt{ibm\_marrakesh}. The cumulants at different final times are shown in Figs.~\ref{fig:cumulants}c and~\ref{fig:cumulants}d, respectively. 
The deviations of the experimental data from the MPS reference solutions at evolution times $T \gtrsim 0.4/J$ are larger for these geometries. We present a detailed analysis of the Trotter errors in a 2D square lattice in the Supplementary Note~2, including the impact of the ordering of terms in the Trotter decomposition in Supplementary Figure~2.
While the IBM quantum platforms provide a natural structure to simulate 1D and heavy-hexagonal lattices, and thus to find efficient circuit decompositions, the square-lattice results are likely to suffer from a greater noise and error impact due to the SWAP overhead from implementing distant-qubit couplings that are not naturally present in the platform. However, we see a clear reduction of the mean defect density with CD at short evolution times. 

We also find qualitative differences for the square lattices compared to the 1D and heavy-hexagonal ones: The skewness $\kappa_3$ has negative values here, indicating that the distribution is asymmetric with a longer tail below the mean. This qualitative difference is seen both in the simulations and in the experimental data. As the equilibrium distributions would be symmetric, a nonzero skewness signals a nonequilibrium situation.

We note that the experimental results are obtained directly from raw data, as we were restricted to the sampling mode of Qiskit to get the complete histograms and perform the cumulant analysis. We found that error suppression methods such as dynamical decoupling did not reduce the noise significantly (see Methods).

\section*{Discussion}

In addition to measuring the predicted plateau of the mean defect density, we gain insight into the nonequilibrium dynamics of rapid quenches by measuring the variance and skewness of the defect density distribution. While in the KZ scaling regime, all cumulants are predicted to exhibit a common universal power-law scaling~\cite{delcampo2018,Cui2020,bando2020probing,king2022coherent}, we find that the dependence of the skewness on the quench time differs from the mean and variance in the fast-quench regime. In particular, in the absence of CD, the skewness has a nonmonotonic dependence on the quench time in all geometries, while the mean and variance show a plateau at $T \lesssim 1/J$ and a monotonic decrease for $T \gtrsim 1/J$. For the ladder and 2D square lattice, the skewness increases as a function of the total evolution time while the mean and variance decrease.
Interestingly, in one dimension and in the heavy-hexagonal geometry, the predicted deviations from a symmetric distribution are smaller by an order of magnitude compared to the ladder and the 2D square lattice, and the distribution is skewed in the opposite direction. In the latter two cases, the rapid quench favors rare events with a defect density below the mean. This nontrivial dependence of the asymmetry of the distribution on geometry is an interesting avenue for further study.

Defect formation is well understood in the limit of slow quenches, as it is described by the QKZM and has been experimentally verified. However, the QKZM's validity is restricted to a low density of kinks. We have demonstrated defect formation for fast quenches away from the adiabatic limit when the defect density of the uncontrolled system saturates to a plateau value independent of the quench time. In this regime, our results establish experimentally that the defect density can be tailored and suppressed by implementing approximate CD controls. This approach opens new avenues for quantum simulation and optimization assisted by CD. In the latter context, the paramagnet-to-spin-glass phase transition is of first-order,  calling for an extension to the fast-quench regime of the dynamics of first-order quantum phase transitions \cite{Rams2019,suzuki2024topological,Sadhasivam2024} and its combination with approximate CD. Our study motivates the use of digital quantum computers for detailed simulations of quantum many-body systems in previously unexplored regimes.

\section*{Methods}

\subsection*{Defect statistics}

We experimentally study the density of kinks in the magnetization $\langle \hat{n}_{\text{def}} \rangle = \frac{1}{2 N_e}\sum_{\langle i,j \rangle}\langle 1 - \hat{Z}_{i} \hat{Z}_{j} \rangle$,
where $N_e$ is the number of edges between qubits. 
We consider various geometries with different numbers of edges: a 1D chain, a three-leg ladder, a 2D heavy-hexagonal lattice, and a 2D square lattice. All geometries have open boundary conditions. 
For the 1D chain, the number of edges is $N_e=N-1$, while for the square lattices, $N_e=N_x(N_y-1) + N_y(N_x-1)$, where $N_x$ is the number of columns and $N_y$ the number of rows. For the heavy-hexagonal geometry with 156 qubits considered here, $N_e = 176$. We note that the definition of the kink density operator $\hat{n}_{\text{def}}$ may count single spin-flip impurities as multiple kinks, depending on the geometry and the position of the spin. For instance, in the one-dimensional chain, an impurity of the type $\ket{\uparrow \uparrow \dots \uparrow \downarrow \uparrow \uparrow \dots}$ would be counted as two kinks. Different definitions of the kink operator were investigated in detail in a recent numerical study of the 1D TFIM~\cite{garcia2024quantumkibblezurek}, where it was shown that for the final Hamiltonian considered here, with no transverse field, the kink definition did not have an effect on the KZ scaling exponent. Note also that for the final Ising Hamiltonian, excitations correspond to kinks in the magnetization and one can relate the kink density with the final-state fidelity (see Supplementary Note~5 and Supplementary Figure~5). On the other hand, for models with continuous symmetries, for instance, the defect density would only give partial information about the final-state fidelity since gapless collective excitations may reduce fidelity even when topological defects are not present.

We analyze the first three cumulants of the defect density distribution. Given a probability distribution of the density of kinks $P(n)=\langle\delta(\hat{n}_{\text{def}}-n)\rangle$, where $\delta(x)$ is the Dirac delta function, its characteristic function is given by the Fourier transform 
$\tilde{P}(\theta)=\mathbb{E}[e^{in\theta}]$~\cite{delcampo2018}. The logarithm of the latter is known as the cumulant generating function, which allows one to define the cumulants $\kappa_q$ through the identity $\log\tilde P(\theta)=\sum_{q=1}^\infty\kappa_q(i\theta)^q/q!$
The first three cumulants $\kappa_1 = \langle \hat{n}_{\text{def}} \rangle$, $\kappa_2 = N_e \left\langle \left( \hat{n}_{\text{def}} - \langle \hat{n}_{\text{def}} \rangle \right)^2 \right\rangle$, and $\kappa_3 = N_e^2 \left\langle \left( \hat{n}_{\text{def}} - \langle \hat{n}_{\text{def}} \rangle \right)^3 \right\rangle$,
respectively, correspond to the mean, variance, and the third central moment. The third central moment dictates the skewness of the distribution. Evaluating them in the initial state $\ket{+}^{\otimes N}$ gives $\kappa_1 = 1/2$, $\kappa_2 = 1/4$, and $\kappa_3 = 0$.

In one dimension, for periodic boundary conditions, the TFIM is exactly solvable through a Jordan-Wigner transformation to the free-fermion basis~\cite{dziarmaga2005dynamics}. To analyze kink formation, the model can be written as an ensemble of independent modes, each with excitation probability $p_k$. As a result, the 
statistics of defect formation can be described in terms of independent Bernoulli trials, where a defect is formed in quasimomentum state $k$ with probability~$p_k$~\cite{delcampo2018}. 
We extend this solution to the fast-quench limit both in the absence and presence of CD and derive the momentum-dependent excitation probability $p_k$ in both cases. The probabilities are obtained by solving numerically the corresponding time-dependent Schr\"odinger equation, as shown in the Supplementary Note~4. 
The leading cumulants up to third order are functions of $p_k$ given by $\kappa_1 = \sum_k p_k$, $\kappa_2 = \sum_k p_k(1-p_k)$, $\kappa_3 = \sum_k p_k(1-p_k)(1-2p_k)$.
In the limit of a sudden quench, $T \to 0$, we recover the simple expressions $\kappa_1 = 1/2$, $\kappa_2 = 1/4$, and $\kappa_3 = 0$ when no CD is applied. In the presence of CD, the cumulants are found as $\kappa_1\approx 0.22$, $\kappa_2\approx0.13$, and $\kappa_3\approx0.04$. We remark that in the presence of CD, a discontinuity exists at $T = 0$: The values of the cumulants in the initial state differ from those in the $T \to 0$ limit. The results in the main text are shown for the defect statistics at the end of the evolution, at~$t = T$. A discussion of the time evolution of the cumulants from $t = 0$ to $t = T$ can be found in the Supplementary Note~6, where Supplementary Figure~6 shows the time-dependent cumulants in the 1D model.

\subsection*{Experiment}
\label{sec:experiment}

The experiments were realized on 156-qubit IBM Heron devices accessed through the cloud using qiskit~\cite{qiskit2024}. We used both IBM Fez and IBM Marrakesh based on availability and calibration at the time of the experiment. To implement the time evolution resulting from $\mathcal{H}(\lambda) = \left(1 - \lambda \right) H_i + \lambda H_f + \dot{\lambda} A_{\lambda}$ on gate-based quantum computers, we used the first-order Suzuki-Trotter decomposition to obtain the digitized time-evolution operator $U(T) = \prod_{m = 0}^M \exp[-i \mathcal{H}(m\delta t) \delta t] + \mathcal{O}(M\delta t^2)$ for $M$ Trotter steps with $\delta t=T/M$ (see~\figlabel{fig:schematic}c). The ordering of the terms $e^{-i\delta t (1 - \lambda) H_i}$, $e^{-i\delta t \lambda H_f}$, and $e^{-i\delta t \dot{\lambda} A_{\lambda}}$ in the circuit does not influence the scaling of the Trotter error with $\delta t$, but for large time steps $\delta t J \not\ll 1$, it may still lead to differences in the results~\cite{tranter2019ordering}. An analysis of these differences in the case of a 2D square lattice is presented in Supplementary Figure~2.

For every geometry, we built the quantum circuits using graph coloring, thereby parallelizing entangling gates and reducing the number of required layers. Similar compression techniques have been applied in a heavy-hexagonal architecture~\cite{kim2023evidence}. Here, we create a conflict graph where each gate is a node. Two nodes are connected if their associated gates overlap. Then, the graph is colored, resulting in chunks of entangling gates that can be applied in parallel. Since the interactions between spins are short-range, greedy optimizers provide optimal solutions for the graph coloring problem. In particular, we use the default solvers of the Networkx library~\cite{hagberg2008exploring}. As an example, the efficient gate ordering for the heavy-hex lattice is shown in the Supplementary Figure~1.

Next, we transpiled the quantum circuits, considering nine different transpilation techniques that may suppress errors and checked which one gave the best performance, i.e. the lowest density of defects relative to the reference 1D values. This test was performed using \texttt{ibm\_fez} with $20000$ shots for a circuit including CD at $T = 0.1/J$ and $dt=0.1/J$ in a $100$-qubit chain with open boundary conditions.

The different transpilation techniques are the combination of three gate scheduling types: as-late-as-possible (ALAP), as-soon-as-possible (ASAP), and none, and using either standard basis gates $\{\text{CZ},X,R_z(\theta),\sqrt{X}\}$, fractional gates~\cite{frac}, and standard basis gates plus dynamical decoupling (DD) with the $X_pX_m$ sequence. In the ASAP schedule, the gates are executed at the earliest available time, which reduces idle times, whereas ALAP delays the gates to the last possible moment, leaving earlier idle periods that can be filled with error-mitigation strategies such as dynamical decoupling. In Table \ref{tab:scheduling_types}, we observe that the best performance resulted from using the standard gates with DD ($X_pX_m$) and scheduling the gates ALAP. Nevertheless, its performance was close to no scheduling and standard gates. Therefore, we used standard gates and no scheduling type.

We employed the default qiskit transpiler with \verb|optimization_level=3| and upon availability, we used the qiskit AI transpiler~\cite{ai_transpiler}. These transpiled circuits were sent to the hardware using the qiskit IBM runtime \verb|SamplerV2| primitive with up to $20000$ shots. The exact gate counts of the transpiled circuits, as well as the device calibrations at the time of experiment, can be found in Ref.~\cite{data_repository}.

\begin{table}[h]
\centering
\caption{The density of defects for different transpilation techniques for a 100-qubit 1D chain on \texttt{ibm\_fez}. The number closest to the exact result is marked in bold font.}
\label{tab:scheduling_types}
\begin{tabular}{c|c|c|c}
\hline
\begin{tabular}[c]{@{}c@{}}Scheduling\\ type\end{tabular} & \begin{tabular}[c]{@{}c@{}}Plain\\ standard\end{tabular} & \begin{tabular}[c]{@{}c@{}}Plain\\ fractional\end{tabular} &  \begin{tabular}[c]{@{}c@{}}Standard\\ + DD ($X_pX_m$)\end{tabular} \\
\hline
\hline
None & 0.268829 & 0.284225  & 0.265135 \\
\hline
ALAP & 0.270566 & 0.285583  & \textbf{0.265124} \\
\hline
ASAP & 0.269721 & 0.280805  & 0.267750 \\
\hline
\end{tabular}
\end{table}

Finally, due to current hardware limitations, the number of time steps in the digitized time evolution was held constant for \(T \geq 0.8/J\) in order to minimize device error (see Supplementary Figure~3). Consequently, the time step size \(\delta t\) was increased. We set the maximum number of Trotter steps to five, which gave no more than $3000$ entangling gates in the studied cases.

\subsection*{Simulations with matrix product states}
For the ladder and two-dimensional geometries, we obtained reference solutions by time-dependent MPS~\cite{Schollwock2011} simulations. We used the time-dependent variational principle (TDVP)~\cite{haegeman2016unifying} with a two-site update, provided in the open-source package \texttt{ITensors.jl~}\cite{ITensor-r0.3}. For the reference lines in~\figlabel{fig:cumulants}b-d, we set the time step to $\delta t = 0.005/J$, the truncation cutoff to $10^{-20}$, and limited the maximum bond dimension to $\chi = 250$. We verified the convergence of the computed observables so that the numerical accuracy is given by the line width in the plots. While the short-time evolution of the systems studied here can be simulated accurately with MPS,
recent work using analog~\cite{king2024computational} and digital-analog~\cite{andersen2024thermalization} quantum simulators has reported evolution times that are inaccessible with classical methods at the same level of precision.

For the two-dimensional geometries, we find that the run times required for the level of accuracy chosen here are up to several hours.
For the heavy-hexagonal lattice, we also record the wall-clock run time of time evolution up to $T = 2.7/J$, corresponding to a single data point in~\figlabel{fig:cumulants}b. To achieve a similar accuracy for $\kappa_1$ as in the experiment, quantified by the difference between the measured $\kappa_1$ and the converged value from the MPS simulation shown in~\figlabel{fig:cumulants}b, we observe a run time of TDVP of around $10$ s. This estimate is obtained for TDVP with a two-site update, run on a MacBook Pro with an Apple M3 chip, 8 cores, and 16 GB of memory.
We note that the run times could potentially be reduced, depending on the geometry, by using different variations of time-evolution algorithms~\cite{yang2020time-dependent,li2024time-dependent,tindall2024efficient,patra2024efficient,tindall2025dynamics}. For increasing system sizes, connectivities, and evolution times, the growth of entanglement will nevertheless eventually prohibit accurate classical simulations, a limitation inherently absent in quantum devices.

\medskip

\noindent{\textbf{\textsf{Data availability}}}\\
The data supporting the findings of this study can be found via figshare~\cite{data_repository}.

\smallskip

\noindent{\textbf{\textsf{Code availability}}}\\
The codes generated and used during the current study
are available from the corresponding author upon reasonable request.
 
 \smallskip

\noindent{\textbf{\textsf{Acknowledgements}}}\\ 
We thank Stefan Woerner for his feedback on the manuscript and Michael Wurster, Sebastian Wagner, and Michael Falkenthal for their help in running the simulations. We acknowledge the use of IBM Quantum services for this work. The views expressed are those of the authors and do not reflect the official policy or position of IBM or the IBM Quantum team. 
This project was supported by the Luxembourg National Research Fund (FNR Grant Nos.\ 17132054 and 16434093). It has also received funding from the QuantERA II Joint Programme and co-funding from the European Union’s Horizon 2020 research and innovation programme.

\smallskip

\noindent{\textbf{\textsf{Author contributions}}}\\
N.N.H., A.G.C. and A.d.C. conceptualized the study. A.-M.V. performed matrix product states simulations and contributed to the analytic derivations and the analysis and interpretation of the results. A.G.C., B.A.B., S.V.R., P.C., and N.N.H. collected the experimental results and contributed to their analysis. B.A.B. performed matrix product states simulations. A.G. and A.d.C. provided the analytic derivation of the defect density cumulants for the one-dimensional model. A.G., A.d.C. and E.S. contributed to the interpretation of the results. All authors participated in writing and reviewing the manuscript.

 \smallskip

\noindent{\textbf{\textsf{Competing interests}}}\\
 The authors declare no competing interests.

\bibliography{reference.bib}
\clearpage

\end{document}

\title{Supplementary Information for:\\Digitized counterdiabatic quantum critical dynamics}

\author{Anne-Maria Visuri$^{\orcidlink{0000-0002-4167-7769}}$}
\email{annemariavisuri@gmail.com}
\affiliation{Kipu Quantum GmbH, Greifswalderstrasse 212, 10405 Berlin, Germany}
\author{Alejandro Gomez Cadavid$^{\orcidlink{0000-0003-3271-4684}}$}
\affiliation{Kipu Quantum GmbH, Greifswalderstrasse 212, 10405 Berlin, Germany}
\affiliation{Department of Physical Chemistry, University of the Basque Country EHU, Apartado 644, 48080 Bilbao, Spain}

\author{Balaganchi A. Bhargava$^{\orcidlink{0009-0006-7595-8776}}$} 
\affiliation{Kipu Quantum GmbH, Greifswalderstrasse 212, 10405 Berlin, Germany}

\author{Sebastián V. Romero$^{\orcidlink{0000-0002-4675-4452}}$}
\affiliation{Kipu Quantum GmbH, Greifswalderstrasse 212, 10405 Berlin, Germany}
\affiliation{Department of Physical Chemistry, University of the Basque Country EHU, Apartado 644, 48080 Bilbao, Spain}

\author{András Grabarits$^{\orcidlink{https://orcid.org/0000-0002-0633-7195}}$}
\affiliation{Department of Physics and Materials Science, University of Luxembourg, L-1511 Luxembourg, Luxembourg}

\author{Pranav Chandarana$^{\orcidlink{https://orcid.org/0000-0002-3890-1862}}$}
\affiliation{Kipu Quantum GmbH, Greifswalderstrasse 212, 10405 Berlin, Germany}
\affiliation{Department of Physical Chemistry, University of the Basque Country EHU, Apartado 644, 48080 Bilbao, Spain}

\author{Enrique Solano$^{\orcidlink{https://orcid.org/0000-0002-8602-1181}}$}
\email{enr.solano@gmail.com}
\affiliation{Kipu Quantum GmbH, Greifswalderstrasse 212, 10405 Berlin, Germany}

\author{Adolfo del Campo$^{\orcidlink{0000-0003-2219-2851}}$}\email{adolfo.delcampo@uni.lu}
\affiliation{Department of Physics and Materials Science, University of Luxembourg, L-1511 Luxembourg, Luxembourg}
\affiliation{Donostia International Physics Center, E-20018 San Sebastian, Spain}

\author{Narendra N. Hegade$^{\orcidlink{0000-0002-9673-2833}}$}
\email{narendrahegade5@gmail.com}
\affiliation{Kipu Quantum GmbH, Greifswalderstrasse 212, 10405 Berlin, Germany}

\begin{abstract}
    In this Supplementary Information, we provide further notes and extended results to support the findings in the main text. We describe in detail how our experiments were conducted on IBM and discuss the impact of different sources of error that might cause deviations from the expected results. We describe the derivation of the first and second-order nested commutator expansions of the adiabatic gauge potential for the transverse-field Ising model. We also provide a step-by-step derivation of the cumulants of the defect number distribution in the fast-quench limit. Finally, for completeness, we discuss the time evolution of the cumulants, which in the main text are only shown at the final time.
\end{abstract}

\maketitle

\tableofcontents

\suppnote{Digital quantum simulation on IBM hardware}
\label{sec:ibm}

An important aspect of preparing and running quantum circuits on hardware is to transpile the required quantum operations according to the corresponding native gate sets provided by the platform. These typically consist of a universal gate set that contains several one-qubit gates and a single two-qubit entangling gate. In our experiments, we used the \texttt{ibm\_fez} and \texttt{ibm\_marrakesh} platforms, gate-based quantum computers composed of 156 superconducting qubits under a heavy-hexagonal coupling map. Their native gate set is composed of
\begin{equation}
    X=\begin{pmatrix}0&1\\1&0\end{pmatrix},\text{ }\sqrt{X}=\frac{1}{2}\begin{pmatrix}1+i & 1-i\\ 1-i& 1+i\end{pmatrix},\text{ } R_z(\theta)=e^{-i\theta Z/2},
\end{equation}
with $\text{CZ}=\diag(1,1,1,-1)$ as an entangling gate. In addition to them, IBM has recently introduced the fractional gates $R_{zz}(\theta)=e^{-i\theta Z_1Z_2/2}$ (with $0<\theta\le\pi/2$) and $R_x(\theta)=e^{-i\theta X/2}$~\cite{frac} in their Heron-based processors.

To mitigate the impact of errors coming from noisy hardware on our results, we employed the \emph{as-late-as-possible} scheduling method as an error suppression technique. It strategically introduces delays in the circuit to maximize the time that each qubit remains in its ground state, which may help improve the outcome fidelity. 
For the heavy-hexagonal lattice, an optimal circuit decomposition can be found relying on the graph coloring theorem. In particular, for each Trotter step, it is possible to reduce all the two-body terms coming from equations~(2) and~(4) in the main text into a depth-three block using the circuit decomposition shown in~\figlabel{fig:graph_coloring}.

\begin{figure}[!tb]
    \centering
    \includegraphics[width=.5\linewidth]{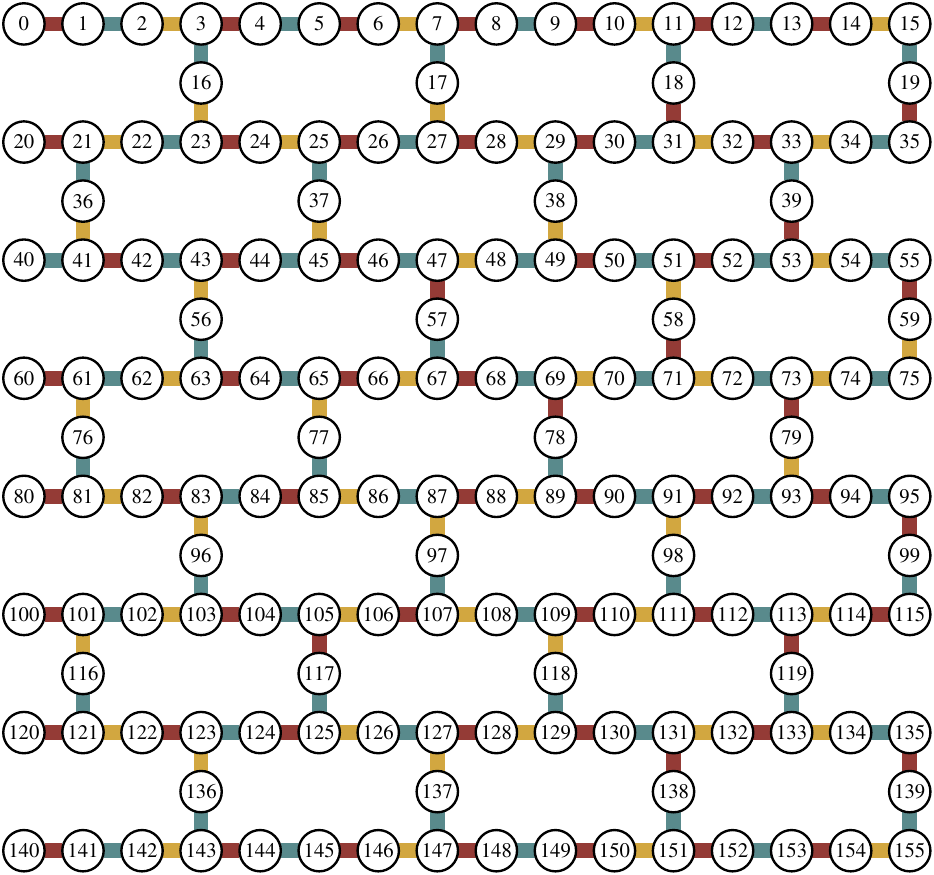}\\\vspace{2mm}
    \includegraphics[width=.5\linewidth]{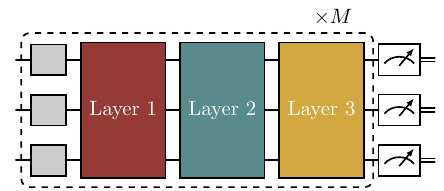}
  \vspace{1mm}
\caption{\textbf{Schematic of the \texttt{ibm\_fez} and \texttt{ibm\_marrakesh} 156-qubit coupling maps.}  Three depth-one layers suffice to realize the two-body interactions between all neighbor pairs for simulating the heavy-hexagonal lattice, finding an optimal circuit decomposition for each Trotter step.}\label{fig:graph_coloring}
\end{figure}

\suppnote{Implementation errors}

\subsection{Trotter errors}
In this section, we analyze various sources of error that can lead to deviations between experimental results and expected outcomes. First, let us consider the Trotter error. We implement a first-order Trotter decomposition of the continuous evolution governed by $\mathcal{H}(\lambda) = H(\lambda) + \dot{\lambda} A_{\lambda}$. For each Trotter step $m = 0, \dots, M$, the quantum circuits realize a product of the two unitaries
\begin{align}\label{eq: addecompo}
    U^m_{ad}(\delta t) &= U^m_f(\delta t) U^m_i(\delta t), \\ 
    \label{eq: cddecompo}
    U^m_{cd}(\delta t) &= e^{-i \delta t/T \cdot 2gJ \alpha_1(\lambda_m)\sum_{\langle i,j \rangle} \hat{Y}_i\hat{Z}_j} e^{-i \delta t/T \cdot 2gJ \alpha_1(\lambda_m)\sum_{\langle i,j \rangle} \hat{Z}_i\hat{Y}_j}.
\end{align}
Here, $\lambda_m = m \delta t$, $\dot{\lambda} = 1/T$,
\begin{equation} \label{eq:initial_final}
    U^m_i(\delta t) = e^{-i \delta t \cdot (1 - \lambda_m) H_i}, \quad \quad 
    U^m_f(\delta t) = e^{-i \delta t \cdot \lambda_m H_f},
\end{equation}
and the initial and final Hamiltonians are $H_i = -g \sum_i\hat{X}_i$, and $H_f = -J \sum_{\langle i,j \rangle}\hat{Z}_i \hat{Z}_j$. The operators $U^m_{ad}(\delta t)$ and $U^m_{cd}(\delta t)$ correspond to digitized annealing and the counterdiabatic (CD) term, respectively. To assess the errors introduced by the first-order Trotter decomposition, we analyze the mean defect density $\kappa_1$ in a square lattice of size $4\times 4$ as a representative case. \figlabel{fig:trotter_error} shows $\kappa_1$ as a function of the total evolution time for various values of the time step $\delta t$. The data indicated by the markers is obtained via matrix product state (MPS) simulations of the quantum circuit implementing the digitized time evolution, performed with the MIMIQ simulator~\cite{QPerfect_MimiQ}. 
As a reference,~\figlabel{fig:trotter_error} shows the results of MPS simulations using the time-dependent variational principle (TDVP)~\cite{haegeman2016unifying,ITensor-r0.3}, with $\delta t = 0.002$, a truncation cutoff of $10^{-20}$, and a maximum bond dimension of $\chi=200$, for which $\kappa_1$ is converged within an accuracy given by the linewidth in the plot. 

\begin{figure*}
    \centering
    \includegraphics[width=\linewidth]{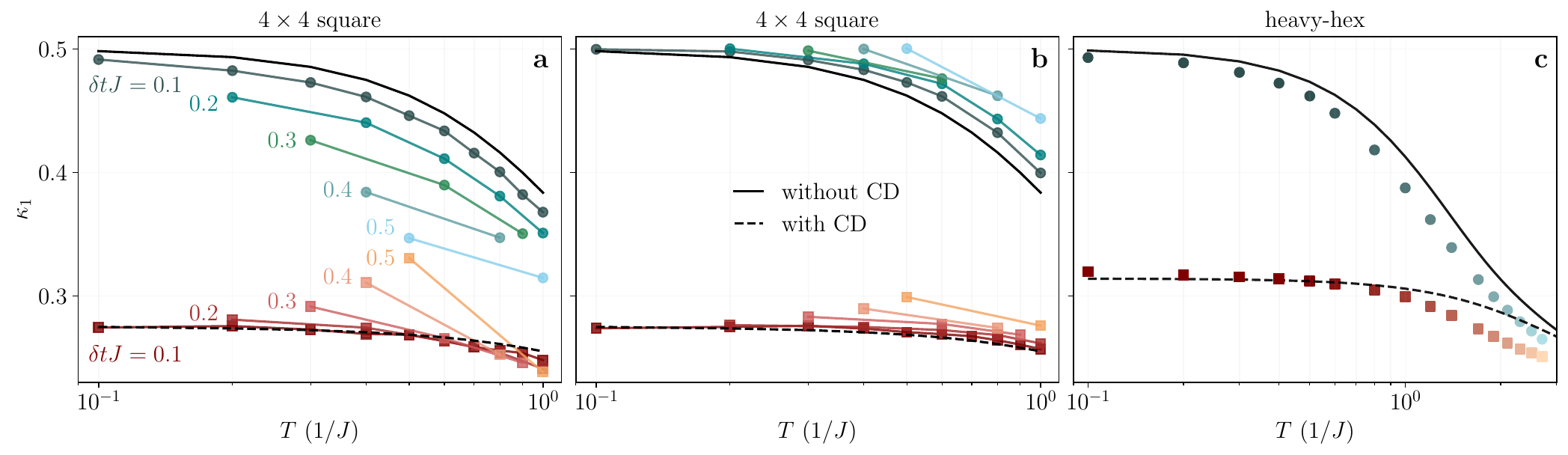}
    \caption{\textbf{Mean density of defects measured at the final evolution time, with and without CD, obtained by MPS simulations of the quantum circuits with MIMIQ~\cite{QPerfect_MimiQ}.} The colored markers denote the circuit simulations that contain Trotter errors, while the solid and dashed black lines show reference MPS simulations of the time evolution with negligible errors. 
    For a square lattice of size $4 \times 4$, we simulate a circuit with different Trotter decompositions: \textbf{a,} $U^m_{ad} = U^m_{f} U^m_{i}$ and \textbf{b,} $U_{ad} = U^m_{i} U^m_{f}$. In both, $U^m_{cd}$ is given by~\eqlabel{eq: cddecompo}. We use varying time steps $\delta t = 0.1/J, 0.2/J, \dots, 0.5/J$ as indicated by the color gradients. The blue-shaded circles correspond to the circuits without CD and the red-shaded squares to those with CD. \textbf{c,} The markers show the circuit simulation results for a 156-qubit heavy-hexagonal lattice with the decomposition $U^m_{ad} = U^m_{f} U^m_{i}$, corresponding to~Fig. 3b in the main text. The time step is varied here in the same way as in~Fig. 3b: $\delta t = 0.1/J$ for the data points up to $T = 0.5/J$ and for larger $T$, the number of steps is fixed to 5 so that $\delta t$ increases. The deviations from the reference solution due to Trotter errors accumulate with increasing $T$.}
     \label{fig:trotter_error}
\end{figure*}

\figlabel{fig:trotter_error} illustrates how the Trotter errors in the quantum circuits lead to deviations in the defect density with respect to the reference values. Although the scaling of the Trotter error with $\delta t$ is independent of the order in which the terms within each time step are applied, the magnitude and sign of the deviations may depend on this ordering (see also Fig.~1c of the main text). In~\figlabel{fig:trotter_error}a, we use the decomposition $U^m_{ad} = U^m_{f} U^m_{i}$ to implement the digitized annealing term, while in ~\figlabel{fig:trotter_error}b, we use the opposite order $U^m_{ad} = U^m_{i} U^m_{f}$. We set $U^m_{cd}$ as in~\eqlabel{eq: cddecompo} in both cases. We find that the magnitude and sign of the deviations differ significantly in these two cases: For digitized annealing without CD (blue-shaded circles), the Trotter errors lead to a reduction of the mean defect density in~\figlabel{fig:trotter_error}a whereas in~\figlabel{fig:trotter_error}b, they lead to an increase with respect to the reference solution. 
For the counterdiabatic evolution (red-shaded squares), in panel~a, the deviations occur in either direction depending on $\delta t$. In panel~b, on the other hand, the deviations are only towards larger defect densities, and the magnitude of the deviations is smaller than in panel~a. These differences have consequences for the experimental data reported in Fig.~3 of the main text: The Trotter decomposition $U^m_{ad} = U^m_{f} U^m_{i}$ is used in all data sets other than the digitized annealing results in  Figs.~3c and d, where we instead use $U^m_{ad} = U^m_{i} U^m_{f}$. We observe that the small deviations of $\kappa_1$ from the reference solution are toward larger values in these experiments, while in Figs.~3b, the experimental $\kappa_1$ obtained by digitized annealing is slightly below the reference line for $T \lesssim 2/J$. 
For the CD evolution, we only observe deviations towards larger defect densities.

In all cases, the circuit simulation results approach the reference lines for decreasing $\delta t$. 
However, increasing $\delta t$ leads to a reduction in the depth of the circuit, since fewer unitary operations are required in the decompositions given by equations~\eqref{eq: cddecompo}--\eqref{eq:initial_final}. 
This reduction is advantageous for experimental implementations, as it significantly mitigates gate errors. Thus, determining an optimal $\delta t$ is crucial to effectively balance Trotter errors and gate errors. In~Figs.~3c and~d of the main text, this balance is carried out using larger $\delta t$ values in larger $T$. This reduces gate errors, but may be a key factor that contributes to the observed increase in the mean defect density as $T$ increases.

We also simulate the quantum circuit implementing the time evolution of the heavy-hexagonal lattice, using the same Trotter steps as in Fig.~3b of the main text, to quantify the Trotter error present in the quantum simulation. The data from a MIMIQ simulation of the quantum circuit is shown in~\figlabel{fig:trotter_error}c together with the MPS reference solution, which is the same as in Fig.~3b. The deviations due to the Trotter error increase as $T$ increases as expected. The results in~\figlabel{fig:trotter_error} suggest that the systematic shift observed in the experimental data for CD evolution, seen in~Fig.~3 of the main text, is not primarily caused by Trotter errors at small $T$ but rather stems from hardware noise. Furthermore, it is important to note that deviations arising from Trotter errors and hardware noise may shift the results in opposite directions, potentially leading to a partial cancellation of errors.

\subsection{Hardware errors}

\begin{figure*}[!tb]
    \centering
    \includegraphics[width=\linewidth]{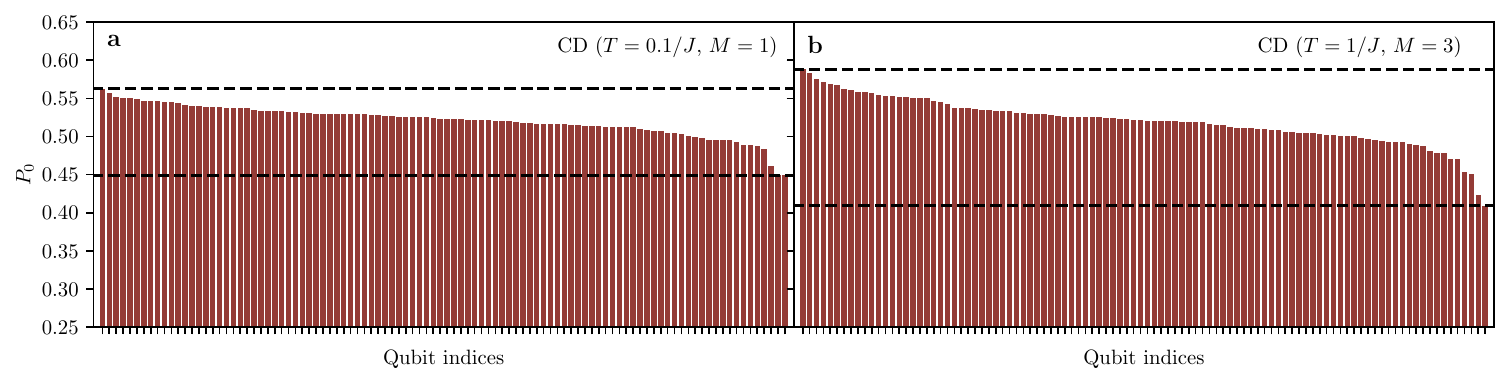}
    \caption{\textbf{Probability per qubit of measuring zero, $P_0$, computed from 20000 samples of the CD-assisted 100-qubit linear chain.} \textbf{a,} Results for a total evolution time $T=0.1/J$ (one Trotter step). \textbf{b,} Results for $T=1/J$ (three Trotter steps). In both panels, the qubit indices are sorted from largest to smallest $P_0$, and the dashed lines indicate the maximum and minimum values of $P_0$.}\label{fig:prob0}
\end{figure*}

Another significant source of error in our results arises from the platform itself, due to gate and measurement errors, short coherence times of the qubits, and other factors. Their impact can be quantified by studying the probability of measuring a zero as an outcome, $P_0$, among all samples taken in our experiments. While for an ideal circuit implementation, we would expect a uniform distribution, where the probability of getting either a zero or a one is equal, qubits with shorter coherence times and bit-flip errors may return different results. In~\figlabel{fig:prob0}, we perform this study for the CD-assisted evolution of a 100-qubit linear chain at times $T=0.1/J$ (one Trotter step) and $T=1/J$ (three Trotter steps), corresponding to results in~Figs.~2a and~3a of the main text. It can be clearly seen that for three Trotter steps, with a larger circuit depth and a larger number of gates, the probability distribution is further from the ideal uniform distribution in comparison to a single step.

\suppnote{Nested-commutator approximation of the counterdiabatic Hamiltonian}

The adiabatic gauge potential (AGP) satisfies the equation $\left[ \partial_{\lambda} H + i[A_{\lambda}, H], H\right] = 0$.
Solving this equation has been shown to be equivalent to minimizing the action $S_{\lambda}(A_{\lambda}) = \text{Tr}\left(G_{\lambda}^{\dagger} G_{\lambda}^{\phantom{\dagger}} \right)$, where
$G_{\lambda}$ is the conserved operator 
\begin{equation}
 G_{\lambda} = \partial_{\lambda} H + i[A_{\lambda}, H].
\end{equation}
This minimization can be performed within a limited subset of operators chosen from the Krylov space obtained by repeatedly applying the Liouvillian $\mathcal{L} = [H, \cdot ]$ onto $\partial_{\lambda} H$~\cite{claeys2019floquet}, resulting in the ansatz of equation~(3) in the main text.
It is convenient to define the operators $\hat{O}_{n} = [H, \hat{O}_{n - 1}]$ for $n = 1, 2, \dots$ and $\hat{O}_0 = \partial_\lambda H$. 
The first-order AGP is then expressed as $A_\lambda^{1} = i  \alpha_1(\lambda)\hat{O}_{1}$.
We use a variational method to find the coefficient $\alpha_{1} \in \mathbb{R}$: 
We minimize the action $S_{\lambda} \left( A_{\lambda}^{1} \right) = S_{1} = \Tr\left(G^{\dagger}_1 G^{\phantom{}}_1 \right)$~\cite{xie2022variantional}, where
    $G_1 = \partial_{\lambda} H + i[ A_{\lambda}^1, H ] = \hat{O}_{0} + \alpha_{1}\hat{O}_{2}$.
The minimization condition requires
\begin{equation}
    \frac{\delta S_{1}}{\delta \alpha_{1}} = 2 \Gamma_{11} + 2 \alpha_{1}\Gamma_{22} = 0,
\end{equation}
where $\Gamma_{ij} = \Tr\left(\hat{O}^{\dagger}_{i}\hat{O}^{\phantom{}}_{j}\right)$, which implies 
\begin{equation} \label{eq:alpha1}
    \alpha_{1} = -\frac{\Gamma_{11}}{\Gamma_{22}}. 
\end{equation}

We can now evaluate $\alpha_1$ for the TFIM. For the Hamiltonian 
\begin{equation}\label{eq:initial_and_final_hamiltonian_suppmat}
H(\lambda) = - \left(1 - \lambda(t) \right) g \sum_{j = 1}^{N} \hat{X}_j - \lambda(t) J \sum_{\langle i,j \rangle} \hat{Z}_i \hat{Z}_j,
\end{equation}
we have
\begin{equation}
\hat{O}_{0} = g\sum_{j=1}^{N}\hat{X}_j - J\sum_{\langle i, j \rangle}\hat{Z}_{i}\hat{Z}_{i}, \quad \quad
    \hat{O}_{1} = -2 i g J \sum_{\langle i, j \rangle}\hat{Y}_{i} \hat{Z}_{j},
\end{equation}
from which we compute $\hat{O}_2$ and obtain the AGP as
\begin{equation}\label{eq:cd_suppmat}
    A_{\lambda}^1 = 2 g J \alpha_1(\lambda) \sum_{\langle i,j \rangle} \left( \hat{Y}_i \hat{Z}_{j} + \hat{Y}_{j} \hat{Z}_i \right).
\end{equation}
When $\alpha_1$ is evaluated for a one-dimensional chain of length $N$, with open boundary conditions, we get
\begin{equation}
    \alpha_{1}(\lambda) = -\frac{(N-1)}{16 g^2 (N-1) (\lambda-1)^2 + 4J^2 [4 (N-1)-3] \lambda^2 }.
\end{equation}
A general formula for~$\alpha_1$ is found in equation~(S20) of ref.~\cite{cadavid2024biasfield}, and results in more complicated expressions that we use for the two-dimensional geometries.

In the following section, we also consider the second-order nested-commutator expansion, where the AGP is given by the expression 
\begin{equation}
    A_{\lambda}^{2} = i\tilde{\alpha}_{1}(\lambda) \hat{O}_{1} + i\tilde{\alpha}_{2}(\lambda) \hat{O}_{3}.
\end{equation}
To solve the coefficients $\tilde{\alpha}_1(\lambda)$ and $\tilde{\alpha}_2(\lambda)$, we minimize the action $S_2 = \mathrm{Tr}\left(G^\dagger_{2} G_{2}\right)$, where $G_{2} = \hat{O}_{0} + \tilde{\alpha}_1 \hat{O}_{2} + \tilde{\alpha}_2 \hat{O}_{4}$. The action can be written as
\begin{align} \label{eq:2nd_order_action}
    S_2 &= \Gamma_{00} + 2 \vec{\alpha}^T
    \begin{pmatrix}
    \Gamma_{02} \\
    \Gamma_{04}
    \end{pmatrix}
    + \vec{\alpha}^T
    \begin{pmatrix}
    \Gamma_{22} &\Gamma_{24} \\
    \Gamma_{24} &\Gamma_{44}
    \end{pmatrix}
    \vec{\alpha},
\end{align}
where $\vec{\alpha}^T = (\tilde{\alpha}_1 \; \tilde{\alpha}_2)$. We have simplified the expression by considering $\tilde{\alpha}_1, \tilde{\alpha}_2 \in \mathbb{R}$ and $\Gamma_{ij} = \Gamma_{ji}$ which will be seen below. 
The stationarity $\delta S_2/\delta \vec{\alpha} = 0$ of the action leads to
\begin{align}
    \begin{pmatrix}
    \Gamma_{22} &\Gamma_{24} \\
    \Gamma_{24} &\Gamma_{44}
    \end{pmatrix}
    \vec{\alpha} = -
    \begin{pmatrix}
    \Gamma_{02} \\
    \Gamma_{04}
    \end{pmatrix}
\end{align}
from which we can solve
\begin{align} \label{eq:alpha1_alpha2}
    \tilde{\alpha}_1 = -\frac{\Gamma_{02} \Gamma_{44} - \Gamma_{04} \Gamma_{24}}{\Gamma_{22} \Gamma_{44} - \Gamma_{24}^2}, \quad \quad
    \tilde{\alpha}_2 = -\frac{\Gamma_{04} \Gamma_{22} - \Gamma_{24} \Gamma_{02}}{\Gamma_{22} \Gamma_{44} - \Gamma_{24}^2}.
\end{align}
In \seclabel{sec:second_order_cd}, we evaluate the $\Gamma_{ij}$ factors explicitly for the 1D TFIM.

\suppnote{Analytical calculation of the cumulants in fast quenches in the 1D chain}
\label{app:analytics}
In this section, we provide an exact solution of the cumulants of the defect density distribution in the $1$D transverse-field Ising model. This extension of ref.~\cite{delcampo2018} from the Kibble-Zurek regime to sudden quenches is discussed in ref.~\cite{grabarits2025drivingquantumphasetransition} for finite-time annealing, while the derivation in the case of counterdiabatic dynamics has not been presented in previous literature.

\subsection{Quench without CD}

We obtain the solution of the $1$D transverse-field Ising model for periodic boundary conditions. We verify, by computing the cumulants numerically using MPS, that the difference due to open boundary conditions is negligible for the system size considered here.
Following a 
Jordan-Wigner transformation to the free-fermion basis (see refs.~\cite{dziarmaga2005dynamics,Berlinsky2019} for a derivation), 
the Hamiltonian $H(\lambda)$ of equation~(1) in the main text is expressed as the direct sum of independent two-level systems (TLSs),
\begin{equation}\label{eq:H_TFIM_momentum}
	H(\lambda) = \sum_{k > 0} \hat{\psi}_k^\dagger H_k \hat{\psi}_k,
\end{equation}
where $H_k = h_{kz} Z + h_{kx} X$,
\begin{equation}\label{eq:H_k}
    h_{kz} = 2\left[ (1-\lambda) g -\lambda J \cos (k a) \right], \quad \quad
    h_{kx} = 2\lambda J \sin (k a),
\end{equation}
and $\lambda = \lambda(t) = t/T$ is the linear scheduling function. In the following, we set the lattice spacing to $a = 1$.
The operator $\hat{\psi}_k:= (\hat{c}_k, \hat{c}_{-k}^\dagger)^T$ is a vector of the annihilation $\hat{c}_k$ and creation $\hat{c}_{-k}^\dagger$ operators for fermions of quasimomentum $\pm k$, where $k=\frac{\pi}{N},\frac{3\pi}{N},\,\dots,\,\left(\pi-\frac{\pi}{N}\right)$. These momenta correspond to the subspace with an even number of fermions, which is consistent with the initial state for even $N$. We denote the Pauli matrices by $X,Y,Z$.
The cumulants are obtained from the time-evolved wave function at the end of the process at $t=T$.

The time evolution of the free-fermion system can be solved for each momentum state independently so that it decomposes into the time evolutions of independent TLSs~\cite{Suzuki2012Quantum}:
\begin{equation}
    i \partial_t
    \begin{pmatrix}
        \varphi_{1,k} \\
        \varphi_{2,k}
    \end{pmatrix}
    = H_k\begin{pmatrix}
        \varphi_{1,k} \\
        \varphi_{2,k}
    \end{pmatrix}.
    \label{eq:time_dependent_se}
\end{equation}
The initial state, ground state of the initial Hamiltonian [equation~(1)], is given by $\varphi_{1,k}=0$ and $\lvert\varphi_{2,k}\rvert=1$ for all $k$, which corresponds to no excited fermion pairs at quasimomenta $(k, -k)$.
With the time-evolved components, one can express the excitation probabilities within each TLS by projecting the time-evolved state at the final time $\ket{\text{FS}_k(T)} = \left(\varphi_{1,k}(T) \: \: \varphi_{2,k}(T) \right)^T$ onto the excited state of the final Hamiltonian,
\begin{equation}
    H_k(T) = 2J
    \begin{pmatrix}
		-\cos k   	&\sin k \\
		\sin k		&\cos k
  \end{pmatrix},\quad
    \ket{\text{ES}_k(T)} = 
    \begin{pmatrix}
        \sin \frac{k}{2} \\
        \cos \frac{k}{2}
    \end{pmatrix}.
\label{eq:excited_state}
\end{equation}
Thus, the excitation probability of the $k$-th state is given by
\begin{equation}
    \label{eq:def_pk}
    p_k = |\bra{\text{ES}_k(T)} \text{FS}_k(T) \rangle|^2 = \left| \sin \frac{k}{2} \, \varphi_{1, k}(T) + \cos \frac{k}{2} \, \varphi_{2, k}(T) \right|^2.
\end{equation}

The statistics of the defect formation can be described in terms of independent Bernoulli trials~\cite{delcampo2018,gomez-ruiz2020full}, where a defect is formed in momentum state $k$ with probability $p_k$. This probability equals the expectation value of the number of defects $\kappa_{1, k}$ in momentum state $k$. The expected density of defects is therefore given by the average of all probabilities, $\kappa_1 = \sum_k \kappa_{1, k}/N = \sum_k p_k/N$. The cumulants satisfy the recursion relation 
\begin{equation}
    \kappa_{q + 1, k} = p_k (1 - p_k) \frac{d \kappa_{q, k}}{d p_k}
\end{equation}
and for finite $N$, we find
\begin{align}
    \kappa_1 = \frac{2}{N} \sum_{k>0} p_k, \quad \quad
    \kappa_2 = \frac{4}{N} \sum_{k>0} p_k(1-p_k), \quad \quad
    \kappa_3 = \frac{8}{N} \sum_{k>0} p_k(1-p_k)(1-2p_k), \label{eq:kappa_sum}
\end{align}
where the summation is over $N$ discrete quasimomenta and the factor $2$ takes into account the sum over $-k$. In the limit $N \to \infty$, we can approximate the sum by an integral $\frac{2}{N} \sum_{k > 0} p_k \to \frac{2}{2 \pi} \int_{0}^{\pi} p(k) dk$, 
\begin{align}
    \kappa_1 = \frac{1}{\pi}\int_0^\pi\mathrm dk\,p(k), \quad \quad
    \kappa_2 = \frac{2}{\pi}\int_0^\pi\mathrm dk\,p(k)[1-p(k)], \quad \quad
    \kappa_3 = \frac{4}{\pi}\int_0^\pi\mathrm dk\,p(k)[1-p(k)][1-2p(k)]. \label{eq:kappa_integral}
\end{align}
We solve the cumulants as functions of the total quench time $T$ by evaluating numerically equations~\eqref{eq:kappa_sum}.
For fast quenches, the leading-order behavior is captured by the sudden quench limit $T \to 0$, in which the final state is identical to the initial one up to a phase factor, leading to the excitation probability 
\begin{equation}
    p_k=\cos^2\frac{k}{2}.
\end{equation}
Substituting this into equations~\eqref{eq:kappa_integral} produces the sudden-quench cumulants $\kappa_1 = 1/2$, $\kappa_2 = 1/4$, and $\kappa_3 = 0$.

\subsection{Quench with first-order CD expansion}

\subsubsection{Variational coefficient of the first-order CD term}
In this section, we derive the fast-quench plateau values for the first three cumulants in the presence of the first-order variational CD. We first derive the expression for the variational coefficient $\alpha_1$ for periodic boundary conditions using the TLS representation.
To obtain $A_\lambda^{1} = i  \alpha_1(\lambda)O_{1}$ in this representation, we express the commutators $O_1$ and $O_2$ as the direct sums of the commutators of each TLS Hamiltonian. With $H_k$ defined in~\eqlabel{eq:H_k} and 
\begin{equation}
    O_{0 k} = \partial_\lambda H_k = -2(g + J \cos (k)) Z + 2J\sin (k) X
\end{equation}
we find the higher-order commutators
$O_{n, k} = [H_k, O_{n - 1, k}]$ for $n = 1, 2, \dots$ as
\begin{equation} \label{eq:o_operators}
\begin{aligned}
O_{2m-1,k} &= \alpha_m \, Y, 
& \alpha_m &= 2^{2m+1}\, i \, gJ \sin(ka)\, \big(h_{kz}^2 + h_{kx}^2\big)^{m-1}, \\[2mm]
O_{2m,k}   &= \beta_m \, \big(h_{kz} X - h_{kx} Z\big), 
& \beta_m  &= 2^{2m+2}\, gJ \sin(ka)\, \big(h_{kz}^2 + h_{kx}^2\big)^{m-1}.
\end{aligned}
\end{equation}
We have exploited the commutation relations of the Pauli matrices, specifically, 
\begin{align}
&[H_k, Y] = -2i \left( h_{kz} X - h_{kx} Z \right), \\[2mm]
&[H_k,\, h_{kz} X - h_{kx} Z] = 2i \, (h_{kz}^2 + h_{kx}^2) \, Y.
\end{align}

The action $S_1$ decomposes into the sum of the TLS actions, $S_{1}=\sum_kS_k=\sum_k\mathrm{Tr}\left(G^\dagger_kG_k\right)$ with $G_k(\lambda) = O_{0 k} + \alpha_1(\lambda) O_{2 k}$. We obtain $\alpha_1$ from \eqlabel{eq:alpha1}, where the $\Gamma_{ij}$ factors can be computed, taking the continuum limit, as: 
\begin{equation} \label{eq:gamma_continuum}
    \Gamma_{ij} = \sum_k \mathrm{Tr} (O_{i k}^\dagger O_{j k}) \approx \frac{1}{\pi} \int_0^{\pi} dk \mathrm{Tr} (O_{i k}^\dagger O_{j k}).
\end{equation}
Exploiting the trace identity for the product of Pauli matrices $\mathrm{Tr}\left( A B \right)=2\delta_{A B}$, where $A, B = X, Y, Z$, we get 
\begin{align}
\Gamma_{11} &= \frac{1}{\pi} \int_0^{\pi} dk \; (8 g J \sin k)^2 \, \mathrm{Tr}(Y^2) = 64 \, g^2 J^2, \\[4mm]
\Gamma_{22} &= \frac{1}{\pi} \int_0^{\pi} dk \; (16 g J \sin k)^2 \, \mathrm{Tr}\!\left[(h_{kz} X - h_{kx} Z)^2\right]
= 1024 \, g^2 J^2 \Big[ \lambda^2 J^2 + (1-\lambda)^2 g^2 \Big].
\end{align}
The minimization condition $\delta S_1/\delta\alpha_1(\lambda)=0$ now leads to
\begin{equation}
    \alpha_1(\lambda) = -\frac{1}{16 [\lambda^2 J^2 + (1-\lambda)^2 g^2]}.
\end{equation}
From this, the CD Hamiltonian 
reads $H^\mathrm{CD} = \sum_{k>0}\,\hat{\psi}^\dagger_kH^\mathrm{CD}_k\,\hat{\psi}_k$, where $H^\mathrm{CD}_k = \dot{\lambda} A_{\lambda, k}^1 = i \dot{\lambda} \alpha_1(\lambda)O_{1k} = h_k Y$ and
\begin{equation}
    h_k = \frac{\dot\lambda\, g J \sin k}{2\left[\lambda^2 J^2 + (1-\lambda)^2 g^2 \right]}.
\end{equation}

\subsubsection{Solution of the time-dependent Schr\"odinger equation}
Within the sudden quench approximation $T\rightarrow 0$, the time evolution is dominated solely by the first-order variational CD term as it is proportional to $1/T$. Thus, the time-dependent Schr\"odinger equation for a given $k$-th TLS reads
\begin{equation} \label{eq:sudden_quench}
\begin{split}    
    i \partial_t
    \begin{pmatrix}
        \varphi_1 \\
        \varphi_2
    \end{pmatrix}
    &= (H_k+h_kY)
    \begin{pmatrix}
        \varphi_1 \\
        \varphi_2
    \end{pmatrix}\approx h_k Y
    \begin{pmatrix}
        \varphi_1 \\
        \varphi_2
    \end{pmatrix}.
    \end{split}
\end{equation}
To compute the exact results in Figs.~2a and~3a of the main text, we integrate this equation numerically using the full Hamiltonian $H_k+h_kY$.
To obtain analytic expressions in the sudden-quench limit $T \to 0$, we neglect the term $H_k$ in the last step. 
Due to the fact that the time dependence appears only as an overall multiplying factor, an exact solution can be obtained by taking the exponential of the integral of $h_k\,Y$. By exploiting the relations $(Y)^{2n}=\mathbb I_2$ and $(Y)^{2n+1}=Y$, the time-evolved state reads
\begin{widetext}
\begin{equation} \label{eq:time-evolved_state}
    \begin{pmatrix}
            \varphi_1(T)\\
            \varphi_2(T)
    \end{pmatrix}=
    e^{-i\int_0^T\mathrm dt h_kY}
    \begin{pmatrix}
            0\\
            1
    \end{pmatrix}
    =\left[\cos\left(\int_0^T\mathrm dt h_k\right)-i\sin\left(\int_0^T\mathrm dt h_k\right)Y\right]
    \begin{pmatrix}
            0\\
            1
    \end{pmatrix}=
    \begin{pmatrix}            
            -\sin\left(\int_0^T\mathrm dt h_k\right)\\
            \cos\left(\int_0^T\mathrm dt h_k\right)
    \end{pmatrix}.
\end{equation}
Due to the derivative $\dot{\lambda}$, we can make a change of variables in the integration, 
\begin{align}
\label{eq:change_of_variables}
    \int_0^T\mathrm dt\,h_k(\lambda(t))
    =\int_0^1\mathrm d\lambda\,\frac{g J}{2\left[\lambda^2 J^2 + (1-\lambda)^2 g^2 \right]}\sin k=\frac{\pi}{4}\sin k,
\end{align}
where we have set $g = J = 1$ for simplicity but the results can be straightforwardly generalized to $g \neq 1$, $J \neq 1$. Since $h_k \propto \dot{\lambda}$, the time integral now only depends on the path in parameter space and not the total duration $T$. More generally, within the sudden-quench approximation $\mathcal{H}(\lambda) = H(\lambda) + \dot{\lambda} A_{\lambda} \approx \dot{\lambda} A_{\lambda}$, the unitary time evolution operator becomes 
\begin{equation}
\label{eq:unitary_operator}
    U(T) = \mathcal{T} e^{-i \int_0^T dt \dot{\lambda}(t) A_{\lambda(t)} } 
    = \mathcal{T}_{\lambda} e^{-i \int_0^1 d\lambda A_{\lambda} },
\end{equation}
where $\mathcal{T}$ denotes time ordering and $\mathcal{T}_{\lambda}$ ordering with respect to $\lambda$ (equivalent when $\lambda(t)$ is monotonic). Consequently, at small total evolution times, the excitation statistics saturate to system- and protocol-dependent limiting values that are independent of $T$.

From equations \ref{eq:time-evolved_state} and \ref{eq:change_of_variables}, we find the final state as
\begin{equation}
    \begin{pmatrix}
            \varphi_1(T)\\
            \varphi_2(T)
    \end{pmatrix}=
    \begin{pmatrix}
            -\sin\left(\frac{\pi}{4}\sin k\right)\\
            \cos\left(\frac{\pi}{4}\sin k\right)
    \end{pmatrix}.
\end{equation}
Knowing the final amplitudes, the excitation probabilities can be expressed as
\begin{equation}\label{eq:p_k_var}
    p_k\approx \left\lvert \sin\frac{k}{2}\varphi_1(T)+\cos\frac{k}{2}\varphi_2(T)\right\rvert^2
\approx \cos^2\frac{k}{2}\cos^2\left(\frac{\pi}{4}\sin k\right)+\sin^2\frac{k}{2}\sin^2\left(\frac{\pi}{4}\sin k\right)-\frac{1}{2}\sin k\sin\left(\frac{\pi}{2}\sin k\right),
\end{equation}
\end{widetext}
given by the projection of the final state onto $\ket{\text{ES}_k(T)}$ of~\eqlabel{eq:excited_state}. The sudden-quench approximation, made above in \eqlabel{eq:sudden_quench}, now results in the cumulants $\kappa_1\approx 0.22$, $\kappa_2\approx0.14$, and $\kappa_3\approx0.04$.

\subsection{Quench with second-order CD expansion}
\label{sec:second_order_cd}

Here, we compute the second-order NC expansion $A_{\lambda}^{(2)} = i \tilde{\alpha}_1(\lambda) O_1 + i \tilde{\alpha}_2(\lambda) O_3(\lambda)$. To solve the coefficients $\alpha_1(\lambda)$ and $\alpha_2(\lambda)$, we minimize the action $S_2 = \sum_k \mathrm{Tr}\left(G^\dagger_{k, 2} G_{k, 2}\right)$, where $G_{k, 2} = O_{0 k} + \tilde{\alpha}_1(\lambda) O_{2 k} + \tilde{\alpha}_2(\lambda) O_{4 k}$. The action is written as in \eqlabel{eq:2nd_order_action}
where the $\Gamma_{ij}$ factors decompose into quasimomentum modes: $\Gamma_{ij} = \sum_k \mathrm{Tr} (O_{i k}^\dagger O_{j k}) \approx \frac{1}{\pi} \int_0^{\pi} dk \mathrm{Tr} (O_{i k}^\dagger O_{j k})$. Using \eqslabel{eq:o_operators} and (\ref{eq:gamma_continuum}), we can evaluate
\begin{align}
    \Gamma_{02} &= 64 J^2 g^2 \\
    \Gamma_{22} &= \Gamma_{04} = 1024 J^2 g^2 
    \left[ (1 - \lambda)^2 g^2 + J^2 \lambda^2 \right] \\
    \Gamma_{24} &= 16384\, g^{2} J^{2} \left[
    g^{4} (1-\lambda)^{4}
    + 3 g^{2} J^{2} (1-\lambda)^{2} \lambda^{2}
    + J^{4} \lambda^{4}
\right] \\
    \Gamma_{44} &= 262144\, g^{2} J^{2} \left[
    g^{6} (1-\lambda)^{6}
    + 6 g^{4} J^{2} (1-\lambda)^{4} \lambda^{2}
    + 6 g^{2} J^{4} (1-\lambda)^{2} \lambda^{4}
    + J^{6} \lambda^{6}
\right].
\end{align}
We plug these expressions into \eqlabel{eq:alpha1_alpha2} to obtain the coefficients
\begin{eqnarray}
\tilde\alpha_1(\lambda) &&= -\frac{g^2 (1 - \lambda)^2 + J^2 \lambda^2}{8 \Bigl( g^4 (1 - \lambda)^4 + g^2 J^2 (1 - \lambda)^2 \lambda^2 + J^4 \lambda^4 \Bigr)}, \\
\tilde\alpha_2(\lambda) &&= \frac{1}{256 \Bigl( g^4 (1 - \lambda)^4 + g^2 J^2 (1 - \lambda)^2 \lambda^2 + J^4 \lambda^4 \Bigr)}.
\end{eqnarray}
The time evolution of the state is given by \eqlabel{eq:sudden_quench}, where $h_k$ is now
\begin{equation}
    h_k = -8 \dot{\lambda} g J \sin k \left[ \tilde{\alpha}_1(\lambda) + 4 \tilde{\alpha}_2(\lambda) (h_{kz}^2 + h_{kx}^2) \right].
\end{equation}
The cumulants can again be computed numerically from \eqslabel{eq:def_pk}, (\ref{eq:kappa_sum}), (\ref{eq:kappa_integral}), and (\ref{eq:sudden_quench}), and they are plotted in \figlabel{fig:thermodynamic_limit}. 

We can evaluate the cumulants in the $T \to 0$ limit using \eqlabel{eq:time-evolved_state}, where
\begin{equation}
    \int_0^T dt h_k(\lambda(t)) = \frac{\pi}{6\sqrt3}\left(3+2\cos k\right)\sin k.
\end{equation}
From \eqlabel{eq:def_pk}, we get the excitation probabilities as
\begin{equation}
    p_k = \sin^2\frac{k}{2}\sin^2\left[\frac{\pi}{6\sqrt3}\left(3+2\cos k\right)\sin k\right]+\cos^2\frac{k}{2}\cos^2\left[\frac{\pi}{6\sqrt3}\left(3+2\cos k\right)\sin k\right]-\frac{1}{2}\sin k\sin\left[\frac{\pi}{3\sqrt3}\left(3+2\cos k\right)\sin k\right].
\end{equation}
Although the expression is somewhat more involved, the first three cumulants can still be obtained using the same integral approximations, Eq.~\eqref{eq:kappa_integral}, as in the previous sections, yielding $\kappa_1 \approx 0.139,\kappa_2 \approx 0.082,\kappa_3 \approx 0.024$. These values match the plateaus in \figlabel{fig:thermodynamic_limit}. As shown in \figlabel{fig:thermodynamic_limit}, the second-order approximation exhibits a stronger suppression of defects than the first-order one. This trend is also reflected in the behavior of the final fidelities, as shown in \figlabel{fig:log_fidelity}a.

\subsection{Cumulants in the thermodynamic limit}

The cumulants of the defect density distribution are shown for a system size $N = 100$ in~Fig.~3 of the main text. For the 1D TFIM, the cumulants can be straightforwardly computed also in the thermodynamic limit $N \to \infty$ by evaluating the integrals of equations~\ref{eq:kappa_integral} numerically with a high resolution in $k$. \figlabel{fig:thermodynamic_limit} shows that the finite-size results for $N = 100$ overlap with the infite-size limit, indicating that finite-size effects at this system size are negligible.

\begin{figure*}[!tb]
    \centering
    \includegraphics[width=\linewidth]{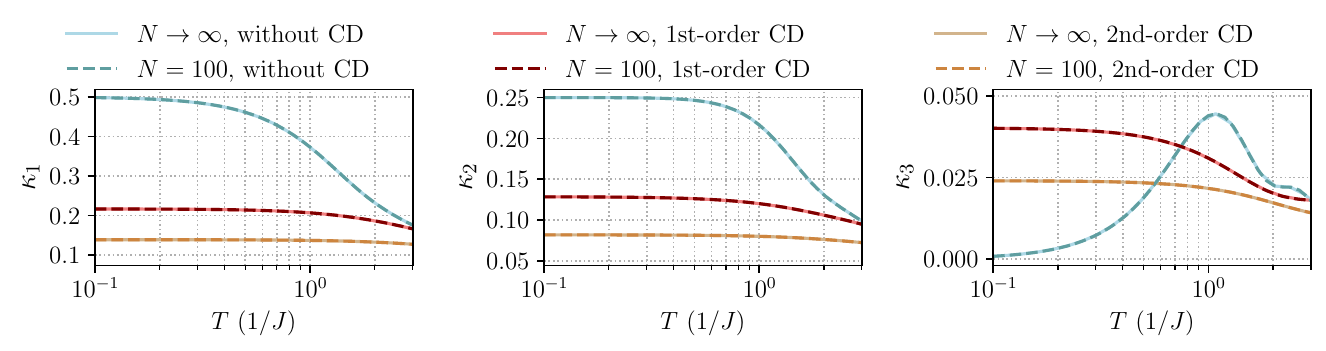}
    \caption{\textbf{Cumulants of the defect density distribution, computed for the 1D TFIM.} The infinite-size limit ($N \to \infty$) and the finite-size results for $N = 100$ overlap, showing that finite-size effects are negligible.}   
    
    \label{fig:thermodynamic_limit}
\end{figure*}

\suppnote{Connection between the defect density and the target-state fidelity}

In the 1D TFIM, the fidelity of the final state with respect to the target ground state can be written in terms of the excitation probabilities $p_k$~\cite{dziarmaga2010dynamics}
\begin{equation}
    f = |\langle \text{FS} | \text{GS} \rangle|^2 = \prod_{k} (1 - p_k),
    \label{eq:fidelity}
\end{equation}
while the density of defects in the final state is $n_{\text{def}} = \frac{1}{N} \sum_k p_k$. We can write
\begin{equation}
   -\ln{f} = -\sum_{k} \ln{(1 - p_k)} = \sum_{k} \left[ p_k + \frac{p_k^2}{2} + \mathcal{O}(p_k^3) \right],
\end{equation}
so that for $p_k \ll 1$, we recover the number of defects as $-\ln f \approx \sum_k p_k = N n_{\text{def}}$. The fidelity therefore scales exponentially with the number of defects in the limit of few defects, $f \approx e^{-N n_{\text{def}}}$. 

While this relation can only be derived for the 1D model, which decomposes into independent TLSs,  \figlabel{fig:log_fidelity} shows that the negative log-fidelity $-\ln(f)/N$ decreases monotonically with the total evolution time $T$ for all the geometries considered in the main text. It is reduced when CD driving is applied, thus the fidelity is enhanced. For the 1D model, the fidelity can be computed exactly for both first- and second order CD approximations, and \figlabel{fig:log_fidelity} shows that using the second-order approximation leads to a higher target-state fidelity than the first-order one.

\begin{figure*}[!tb]
    \centering
    \includegraphics[width=\linewidth]{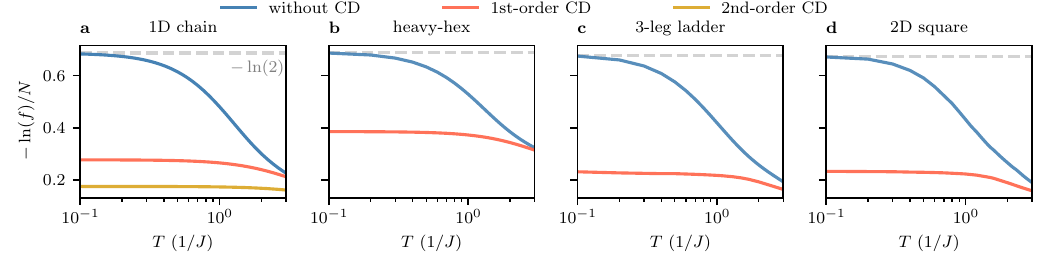}
    \caption{\textbf{The average negative log-fidelity $-\ln(f)/N$ decreases monotonically with total time}, similar to the mean defect density in Fig.~3 of the main text. The lattice geometries are the same as in Fig.~3. Without CD, the $T \to 0$ limit coincides with the value obtained for the initial state, indicated by the gray dashed line. 
    }
    \label{fig:log_fidelity}
\end{figure*}

The negative log-fidelity has a qualitatively similar behavior to the mean defect density shown in \figlabel{fig:thermodynamic_limit}a and in Fig.~3 of the main text. Without CD, the system is far from the limit $p_k \ll 1$, and $-\ln(f)/N > \kappa_1$. The approximate CD driving reduces defects and $-\ln(f)/N$ approaches $\kappa_1$. The $T \to 0$ limit coincides with $-\ln(f)/N$ of the initial state with respect to the final state, $|\bra{\psi_0} \psi_{f} \rangle|^2 = 1/2^{N - 1}$, where $\ket{\psi_0} = \ket{+}^{\otimes N} = \left[ (\ket{0} + \ket{1})/\sqrt{2} \right]^{\otimes N}$ and  $\ket{\psi_f} = (\ket{0 0 \dots 0} + \ket{1 1 \dots 1})/\sqrt{2}$. The 1D results in~\figlabel{fig:log_fidelity}a are obtained using~\eqslabel{eq:time_dependent_se}, (\ref{eq:def_pk}), (\ref{eq:sudden_quench}), and~(\ref{eq:fidelity}), while the results in~\figlabel{fig:log_fidelity}b--d are obtained from MPS simulations.

\suppnote{Cumulants as functions of instantaneous time}

The main text discusses the defect statistics at the final time. For completeness, we investigate here how the defect density cumulants evolve in time, in particular the features that arise when crossing the phase transition. The defect density defined in the main text
\begin{equation}
    \hat{n}_{\text{def}} = \frac{1}{2N_e} \sum_{\langle i, j \rangle}(1 - \hat{Z}_i \hat{Z}_j)
    \label{eq:ndef}
\end{equation}
corresponds to kinks in the magnetization in position basis. These are topological excitations with respect to the final ferromagnetic ground state where all spins are aligned. The cumulants of the kink density are obtained as 
\begin{equation} \label{eq:kink_cumulants}
\kappa_1 = \langle \hat{n}_{\text{def}} \rangle, \quad
\kappa_2 = N_e \left\langle \left( \hat{n}_{\text{def}} - \langle \hat{n}_{\text{def}} \rangle \right)^2 \right\rangle, \quad 
\kappa_3 = N_e^2 \left\langle \left( \hat{n}_{\text{def}} - \langle \hat{n}_{\text{def}} \rangle \right)^3 \right\rangle. 
\end{equation}
Evaluating them in the initial state $\ket{+}^{\otimes N} = \left[ (\ket{0} + \ket{1})/\sqrt{2} \right]^{\otimes N}$ gives $\kappa_1 = 1/2$, $\kappa_2 = 1/4$, and $\kappa_3 = 0$, and the final-state values are shown as functions of~$T$ in Fig.~3 of the main text.

Section~\ref{app:analytics}, on the other hand, introduces the cumulants of the excitation density, obtained through the excitation probabilities $p_k$ in quasimomentum basis in equations~\eqref{eq:kappa_sum}. These are based on excitation probabilities with respect to the instantaneous ground state and coincide with the cumulants of the kink density at the final time $t = T$, where the ground state has all spins aligned. However, at $t < T$, the density of kinks differs from the density of excitations. In the following, we denote the cumulants of the excitation density at $t<T$ by $K_q$, with $q=1,2,3$, corresponding to the mean, variance, and third central moment, respectively. 

\begin{figure*}[!tb]
    \centering
    \includegraphics[width=0.65\linewidth]{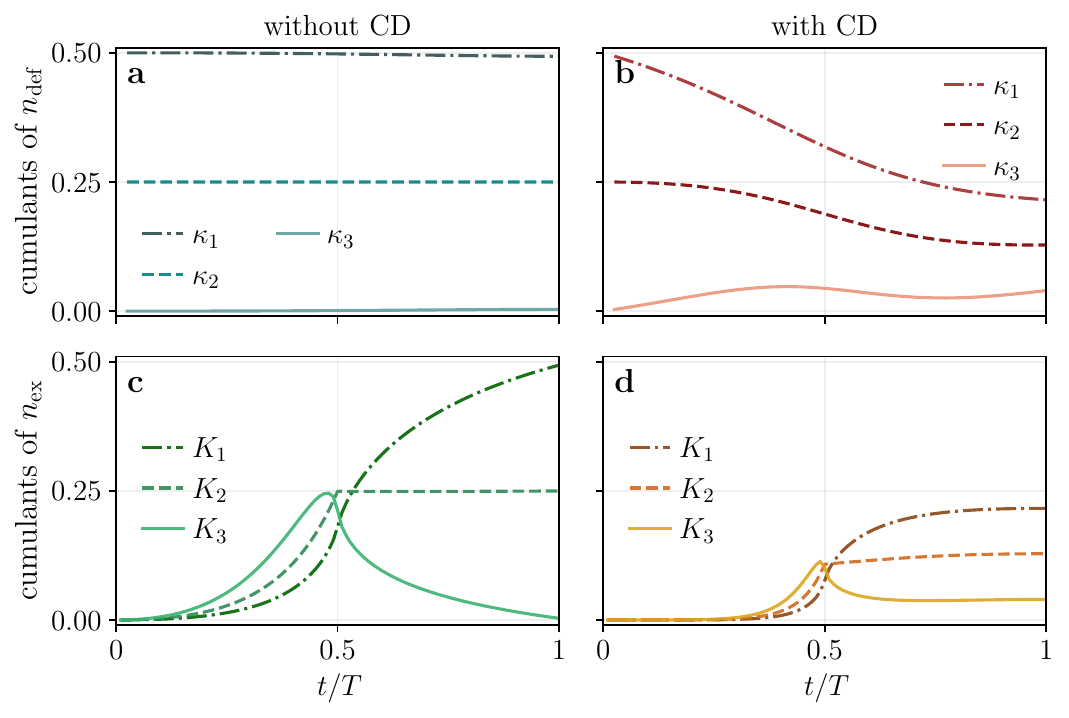}
    \caption{\textbf{The three first cumulants of the distribution of \textbf{a, b,} kinks and \textbf{c, d,} excitations as functions of the unitless time $t/T$.} Here, we consider a one-dimensional lattice of length $N = 100$ and set $T = 0.2/J$. \textbf{a, b} The kink density cumulants defined in position basis, given by equations~\eqref{eq:ndef} and~\eqref{eq:kink_cumulants}, differ from \textbf{c, d,} the cumulants of the excitation density $n_{\text{ex}}$ defined in quasimomentum basis [\eqlabel{eq:kappa_integral}] 
    at intermediate times $t < T$. We use the notation $K_q$ for the cumulants of the excitation density at $t < T$, and their definition coincides with $\kappa_q$ at the final time $t = T$. The kink density cumulants in panels \textbf{a} and \textbf{b} are obtained from an MPS simulation. 
    }
    \label{fig:instantaneous_cumulants}
\end{figure*}

The cumulants of the kink density and excitation density are shown as a function of the unitless time $t/T$ in~\figlabel{fig:instantaneous_cumulants} with and without CD. We focus here on the fast-quench regime with total time $T = 0.2/J$. For such short evolution times, the kink density cumulants remain close to their initial values when no CD is applied, as seen in~\figlabel{fig:instantaneous_cumulants}a. In the presence of CD, the kink density and its variance are reduced, and the third central moment slightly increases to its final value shown in~Fig.~3a in the main text. While the kink density cumulants do not have any distinguishing features at the phase transition point $t/T = 0.5$, the cumulants of the excitation density in~\figlabel{fig:instantaneous_cumulants}a and~\figlabel{fig:instantaneous_cumulants}b display a nonanalytic behavior. As the initial state is the ground state of $H_i$ and there are no excitations, the cumulants $K_q$ are zero at $t = 0$. The mean density of excitations $K_1$ grows monotonically, with an inflection point close to where the phase transition occurs. The variance $K_2$ increases up to the critical point and stays nearly constant at $t/T > 0.5$, with a discontinuous derivative at $t/T = 0.5$, while the third central moment $K_3$ has a peak at $t/T = 0.5$. The excitation statistics, therefore, show signatures of the phase transition that are not observed by measuring the kinks in the spin alignment. The same features occur in both finite-time annealing and counterdiabatic evolution, but $K_1$ and $K_2$ are suppressed by CD while $K_3$ increases slightly.

\par\bigskip
\noindent{\large\bfseries Supplementary References\par}
\medskip

\bibliography{reference}
\clearpage